\documentclass[11pt]{article}
\usepackage{jheppubmod}
\pdfoutput=1
\usepackage{framed}
\usepackage{psfrag}
\usepackage{array}
\usepackage{bm, bbm, bbold}
\usepackage{amssymb}
\usepackage{amsmath}
\usepackage{amsthm}
\usepackage{braket}
\usepackage{hyperref}
\usepackage{graphicx}
\usepackage[labelsep=quad]{subcaption}
\usepackage[punctsep]{collref}
\collectsep[]{;}
\usepackage{epsfig}
\usepackage{wrapfig}
\usepackage{tikz}
\usetikzlibrary{arrows,plotmarks,positioning,calc}
\usepackage{environ}
\NewEnviron{eqn}{
\begin{align}
\begin{split}
  \BODY
\end{split}
\end{align}
}

\NewEnviron{eqn*}{
\begin{align*}
\begin{split}
  \BODY
\end{split}
\end{align*}
}

\usepackage[
    natbib=true,
    style=numeric,
    sorting=none, 
    maxbibnames=99
]{biblatex}
\def\pb[#1,#2]{\{#1, #2\}}
\def\deb[#1,#2]{[#1,#2]_{\text{D.B.}}}

\def\tr{{\rm Tr}}

\def\ep{\epsilon}
\def\a{\alpha}
\def\G{\Gamma}

\def\Or[#1]{{\text{O}}\left({#1}\right)}
\def\dotl[#1,#2]{\left\langle #1,\, #2 \right\rangle}
\def\dotlb[#1,#2]{\left\langle #1,\, #2 \right\rangle}
\def\dotlm[#1,#2]{\left[ #1,\, #2 \right]}
\def\dotp[#1,#2]{(\vect{#1} \cdot\vect{#2})}
\def\aff[#1,#2]{\hat{#1}(#2)}

\def\n4sym{{\cal N}=4 SYM}
\def\>{\rangle}
\def\<{\langle}

\def\projsho[#1]{{\cal P}^{\text{sho}}_{#1}}
\def\transsho[#1,#2]{{\cal T}^{\text{sho}}_{#1,#2}}
\def\wedgeeight[#1,#2,#3]{\{(#1),#2,#3\}}
\def\ads[#1]{$\text{AdS}_{#1}$}

\newcommand{\be}{\begin{equation}}
\newcommand{\ee}{\end{equation}}
\newcommand{\ba}{\begin{align}}
\newcommand{\ea}{\end{align}}
\newcommand{\bs}{\begin{split}}
\def\sess\end{split}
\newcommand{\vect}[1]{{\boldsymbol{#1}}}

\def \bea {\begin{eqnarray}}
\def \eea {\end{eqnarray}}
\def \bea* {\begin{eqnarray*}}
\def \eea* {\end{eqnarray*}}

\def \bes {\begin{equation*}}
\def \ees {\end{equation*}}

\def \b  {\beta}

\def\p{\partial}
\newcommand{\tf}{\mbox{\footnotesize tf}}

\def \scrip{{\cal I}^{+}}
\def \scrim{{\cal I}^{-}}

\def\alcut[#1]{{\cal A}_{#1, \epsilon}}
\def\alseg[#1,#2]{{\cal B}_{#1, #2}}

\def\supcharge[#1]{\{#1\}}

\def\projsupeig[#1]{{\cal P}_{{\ell, m}}[{#1}]}
\def\transop[#1, #2]{T_{\{#1\}, \{#2\}}}
\def\supket[#1]{|\{#1\} \rangle}
\def\supbra[#1]{\langle \{#1\} | }

\def\rsop[#1]{X_{#1}}

\def\projlow[#1,#2]{P_{{#1}<{#2}}}
\title{%On Symmetries of Gravitational Scattering in Higher Dimensions
The Asymptotic Structure of Gravity in Higher Even Dimensions
}
\newcommand{\fourshear}{C_{ab}^{(4D)}}
\newcommand{\intI}{\int_{\scrip}}
\date{\today}

\author[a]{Chandramouli Chowdhury,}
\author[a, b, c]{Ruchira Mishra} 
\author[a, d]{and Siddharth G. Prabhu}
\affiliation[a]{International Centre for Theoretical Sciences, Tata Institute of Fundamental Research, Shivakote, Bengaluru 560089, India.}
\affiliation[b]{Department of Physical Sciences, Indian Institute of Science Education and Research - Mohali,
Knowledge City, Sector 81, SAS Nagar, Punjab 140306, India}
\affiliation[c]{Department of Physics, University of Chicago, Chicago IL 60637-1433, USA}
\affiliation[d]{Department of Theoretical Physics, Tata Institute of Fundamental Research, Homi Bhabha Rd,
Mumbai 400005, India}
\emailAdd{chandramouli.chowdhury@icts.res.in}
\emailAdd{ruchiramishra@uchicago.edu}
\emailAdd{siddharth@theory.tifr.res.in}

\abstract{	We investigate the notion of asymptotic symmetries in  classical gravity in higher even dimensions, with $D = 6$ space-time dimensions as the prototype. Unlike in four dimensions, certain non-linearities persist which necessitates the complete non-linear analysis we undertake. We show that the free data is parametrized by a pair of symmetric trace-free tensors at future (past)  null infinity. This involves a redefinition of the radiative field. We define a symplectic structure generating the radiative phase space at ${\cal I}^{\pm}$ with appropriate boundary conditions which are preserved by the action of supertranslations. We derive the charge associated with super-translation vector fields  and show that it matches with the charge derived using the equations of motion in the full non-linear theory. We elaborate on the precise relationship between the super-translation charge, soft theorem and the ``gravitational memory'' in six space-time dimensions, thus providing the first example of an infrared triangle in non-linear gravity beyond four dimensions.}

\allowdisplaybreaks

\begin{document}
\maketitle
%\newpage
%%%%%%%%%%%%%%%%%%%%%%%%%%%
\section{Introduction}
%%%%%%%%%%%%%%%%%%%%%%%%%%% 
 Starting with the seminal work of Bondi-Metzner-Sachs (BMS) and van der Burg in the 1960's, we have learnt a great deal about the asymptotic structure of gravity in four dimensions \cite{Bondi:1962px, Sachs:1962wk, Ashtekar:1981sf, Ashtekar:1981bq}. While originally studied because of its inherent importance in understanding the nature of gravity, the study of asymptotic symmetries is crucial in formulating the idea of holography in asymptotically flat spacetimes. In recent years, the work by Strominger and collaborators \cite{Strominger:2013jfa, He:2014laa, Strominger:2017zoo}  has been of central importance  for reviving the interest in this subject, and also for uncovering a remarkable connection between asymptotic symmetries, soft theorems and the memory effect.
 
 In this paper we investigate the symmetries of asymptotically flat spacetimes near Null Infinity\footnote{The interested reader can refer to \cite{Campiglia:2015kxa, Campiglia:2017mua, Campiglia:2018dyi, Prabhu:2018gzs, Prabhu:2019fsp} for a discussion of asymptotic symmetries near timelike and spacelike infinities.}.  %There has also been a lot of work in the study of asymptotic symmetries in asymptotically de Sitter and Anti de Sitter spacetimes, and we refer the interested reader to \cite{Ferreira:2016hee, Compere:2020lrt, Ashtekar:1999jx, Anninos:2010zf, Banerjee:2020dww}.
 %The symmetry group preserving the asymptotic structure of Null infinity shall be referred to as the asymptotic symmetry group. 
 Ever since the work of BMS, it has been known that the asymptotic symmetry group of flat spacetime in four dimensions is larger than the Poincare group and is known as the BMS group. This contains additional symmetries known as {\it supertranslations}\footnote{Supertranslations are ``angle dependent translations'', where the angle is defined with respect to the sphere at null infinity (also known as the Celestial sphere). We will define this concretely in the main text.} along with the usual Poincare transformations. These asymptotic symmetries have been extensively studied in four dimensions \cite{Strominger:2017zoo, Melas_2017} and there has been some progress in understanding them in the linearized gravitational theory in higher even dimensions as well \cite{Kapec:2015vwa, Aggarwal:2018ilg}\footnote{The behaviour of solutions in odd dimensions differs greatly from that of even dimensions and shows non-integer fall offs. In this work we only consider even dimensional spacetimes.}. However, there are crucial differences between the situations in four and higher dimensions. In four dimensions, supertranslations and radiative degrees of freedom are described by the same field which constitutes the free data. In higher dimensions, these are accounted for by two different fields, occuring at two different fallofs in the asymptotic expansion.  While in four dimensions the theory is effectively free as one goes to null infinity and therefore a linearized analysis is often sufficient, in higher dimensions, certain non-linearities persist at null infinity\footnote{The non-linear action of the supertranslations on the free data was also pointed out in \cite{Aggarwal:2018ilg}.}.  This leads to a tension with the boundary conditions needed for defining the symplectic stucture. As we discuss in the main text, this tension also exists while studying the linearized theory about a non-trivial background.
 
%We study the generators of supertranslations in the full non-linear theory and compute their action on the free data.
In this work, we resolve this issue by first redefining the fields that contain the radiative data. This redefinition does not affect the News tensor characterizing the radiation. Using this redefinition we find the Symplectic structure of the resulting phase space and impose boundary conditions that are consistent with the action of supertranslations. This allows us to compute the Noether charge corresponding to supertranslations. We also find this charge by an explicit evaluation of the equations of motion of the full non-linear theory after first expressing it as an integral of the Bondi mass. We show that these two expressions for the Noether charge agree with each other. 

The Noether charge corresponding to supertranslation symmetry gives a description of the classically conserved quantities. In the quantum regime, these classical conservation laws manifest as the Ward identities, which help us understand the Weinberg soft theorems. The connections between asymptotic symmetries and soft theorems are well established in the literature for four dimensions and in higher dimensional linearized theory \cite{Strominger:2017zoo, Kapec:2015vwa, Aggarwal:2018ilg}. In this paper, we show that this connection persists in the non-linear theory by demonstrating that it retains the same structure as in the linearized theory when described in the redefined variables. 

Along with soft theorems, asymptotic symmetries have also been important for understanding the {\it memory effect}. Such a connection has been well understood in four dimensions and in the linearized theory in higher-dimensions \cite{Strominger:2017zoo, Strominger:2014pwa, Hollands:2016oma, Satishchandran:2019pyc}. Together, asymptotic symmetries, soft theorems and the memory effect complete the {\it Infrared Triangle}. The triangle represents a set of mathematical operations which relate the three seemingly different concepts. In this paper we explain how one can derive the memory effect in the non-linear theory which then leads us to a generalization of the Infrared triangle in higher-dimensions. This provides the first example of the Infrared triangle in the full non-linear theory in higher-dimensions. 

We end with some comments on the implication of our results in the quantum theory in section \ref{sec:quantumdisc}.

%%%%%%%%%%%%%%%%%%%%%%%%%%%
\section{Supertranslations in higher dimensions}\label{sec:STshigher}
%%%%%%%%%%%%%%%%%%%%%%%%%%%
\subsection{Metric Conventions and Gauge Choice}
Although our main aim is to analyze the implications of allowing supertranslations in higher dimensional gravity, for concreteness we work in six dimensions. However, many of the qualitative features should not change for higher even dimensions. In contrast to earlier works \cite{Kapec:2015vwa, Aggarwal:2018ilg} we work in the full non-linear theory. As we go along, we highlight the key differences  between asymptotic symmetries in linearized gravity and our results.  We denote the retarded Bondi cooridnates by $(u, r, z^a)$ with the small Latin indices denoting the coordinates on the Celestial sphere $\mathbb S^4$ and the Greek indices denote the spacetime coordinates, $\mu \in (u, r, z^a)$. Although our analysis is restricted to six spacetime dimensions, several results are independent of the dimensionality and we will highlight them explicitly. We work in the unit of $c = \hbar =  8\pi G_N = 1$ which is the same as \cite{Aggarwal:2018ilg}.  

In the Bondi gauge the (asymptotically) flat metric can be written as \cite{Bondi:1962px, Sachs:1962wk},
\be\label{metric101}
ds^2 = e^{2\b} M du^2 - 2 e^{2\b} du dr + g_{ab}(dz^a - U^a du)(dz^b - U^b du). 
\ee
All quantities in the metric can be a function of $(u, r, z)$. The indices of $U^a$ are raised and lowered using $g_{ab}$, i.e, $U_a = g_{ab} U^b$ and $U^2 = g^{ab} U_a U_b$, where the inverse $g^{ab}$ is defined as $g^{ac} g_{cb} = \delta^a_b$. The inverse metric is then written in the following form,
\be
g_{\mu\nu} = \left( \begin{array}{ccc}
               M e^{2 \b} + U^2 & - e^{2\b} & - U_a \\
               - e^{2\b} & 0 & 0 \\
               - U_b & 0 & g_{ab}
              \end{array}
\right), \qquad
g^{\mu\nu} = \left( \begin{array}{ccc}
               0 & - e^{-2\b} & 0\\
               - e^{-2\b} & -Me^{-2\b} & -U^a e^{-2\b} \\
               0 & -U^b e^{-2\b}  & g^{ab}
              \end{array}
\right).
\label{bigmetric}
\ee
The Bondi gauge condition also places a constraint on the determinant of the metric $g_{ab}$,
\be
\det\left(\frac{g_{ab}}{r^2}\right) = \det(\gamma_{ab})
\label{eq:detcondition}
\ee
where $\gamma_{ab}$ is the metric on the 4-sphere. As shown in appendix \ref{app:inversetrace}, the determinant condition helps us fix the traces of $g_{ab}^{(n)}$. As we are interested in the analysis near null infinity, we will be mostly interested in the large$-r$ limit of the metric components. Depending on how the large$-r$ limit is taken we can either end up near $\scrip$ or $\scrim$ (the future/past null infinity) but the basic construction remains the same in either case, and therefore we shall stick to working near $\scrip$. 

We will need the fall off conditions near $\mathcal I^+$ which are given as, 
\begin{eqn}
&\b = \sum_{n = 2}^\infty \frac{\b^{(n)}(u, z)}{r^n}, \qquad M = \sum_{n = 0}^{\infty} \frac{M^{(n)}(u, z)}{r^n}, \qquad U_{a} = \sum_{n = 0}^\infty \frac{U_a^{(n)}(u, z)}{r^n}, \\
&g_{ab} = r^2 \gamma_{ab} + \sum_{n= -1}^\infty \frac{g_{ab}^{(n)}}{r^n}. 
\label{eq:falloffs}
\end{eqn}
These fall off conditions have been motivated in the literature, see eg. \cite{Kapec:2015vwa, Aggarwal:2018ilg, Strominger:2017zoo}. Further, these conditions imply that the components of the Ricci tensor fall off as 
\begin{eqn}\label{RicciFall}
 &R_{uu} = O(r^{-4}), \qquad R_{ur} = O(r^{-5}), \qquad R_{ua} = O(r^{-4}), \\
 &R_{rr} = O(r^{-6}), \qquad R_{ra} = O(r^{-5}), \qquad R_{ab} = O(r^{-4}).
\end{eqn}
 The fall off conditions are chosen to ensure the finiteness of energy flux and other physical observables when we couple matter fields with gravity. Note that without including external matter fields we do not need to specify any fall off condition for the Ricci Tensor (as they become zero to arbitrary orders in $r$ when evaluated on-shell), however the addition of matter fields necessitates the fall off conditions mentioned above and therefore also constrain the fall off conditions of the matter stress tensor which are equivalent to the fall off for $R_{\mu\nu}$.

%The radiative data is associated to coefficients of the $ g_{ab}(u, r, z)$ in the $1/r$ expansion in all even dimensions. 
For convenience we denote some of the important components of $g_{ab}(u, z)^{(n)}$ using the following notations
\be\label{gabcomp}
g_{ab}^{(-1)} \equiv C_{ab}, \quad 
g_{ab}^{(0)} \equiv D_{ab}, \quad 
g_{ab}^{(1)} \equiv E_{ab}, \quad 
g_{ab}^{(2)} \equiv F_{ab}.
\ee
The indices of $U_{a}^{(n)}$ and $g_{ab}^{(n)}$ are lowered and raised using the metric $\gamma_{ab}$. The determinant condition eq.\eqref{eq:detcondition} fixes the traces of $g_{ab}^{(n)}$, for example,
\begin{subequations}
\begin{align}
C_a^a &= 0~,\\
D_a^a &= \frac{1}{2} C^{ab} C_{ab}~,\\
E^{a}_a &= C\cdot D - \frac{1}{3} C^3~,\\
F^a_a &= C^{ab} E_{ab} + \frac{1}{2} D^{ab} D_{ab} - C^{am} C_{bm} D^b_a + \frac{1}{4} C^a_b C^b_c C^c_d C^d_a~.
\end{align}
\label{traceresults}
\end{subequations}
A detailed derivation of this is given in appendix \ref{app:inversetrace}. It is easily seen that in the linearized limit (which means that non-linear terms involving $g_{ab}^{(n \geq -1)}$ are neglected), the traces of $D_{ab}, \ E_{ab}, \ F_{ab}$ just become $0$. In fact in the linearized limit, the trace of every $g^{(n)}_{ab} = 0 \ \forall \ n \geq -1$ as shown in \cite{Aggarwal:2018ilg}. One can repeat the above anlaysis at $\scrim$. The interested reader is encouraged to look at \cite{Kapec:2015vwa} for more details.

For the sake of convenience, we decompactify the celestial sphere. Thus $\gamma_{ab}$ is simply the flat Euclidean metric $\delta_{ab}$. As the leading order term $M^{(0)}$ in the expansion of $g_{uu}$ is related to the curvature scalar of $\gamma_{ab}$, it vanishes in the case of celestial plane, i.e, $M^{(0)} = 0$ when $\gamma_{ab} = \delta_{ab}$. We will often be referring to $\gamma_{ab} = \delta_{ab}$ as the {\it flat sphere}. This choice of coordinates has been adopted in many papers including \cite{He:2019jjk, Dumitrescu:2015fej, He:2019pll, Kapec:2017gsg} and the explicit coordinate transformation from the round sphere to the flat sphere is provided in the references above. The simplification will not alter any of the results, as at the end of the computation we could always ``covariantize'' the result and revert back to $\gamma_{ab}$ as the sphere metric\footnote{We note that the flat metric on the celestial sphere has already been used in \cite{He:2019jjk} to compute the conserved charges in electromagnetism in higher dimensions, which we independently reproduce from a calculation using the symplectic form in appendix \ref{app:QEDform}. This has also been used in the context of gravity in \cite{Kapec:2017gsg} and we refer the reader to that paper for further details on these coordinates. It can also be seen from our computations that we get the expected results in linearized gravity by working with the flat sphere metric.}.

%%%%%%%%%%%%%%%%%%%%%%%%%%%
\subsection{Free data}
In the previous section we parametrized the space of asymptotically flat spacetimes in the Bondi gauge. In this section we analyze the radiative phase space of the theory at future null infinity. In $D = 4$, the radiative phase space of General Relativity is coordinatized by the shear $\fourshear(u, z)$. It is a linear space of the unconstrained free data such that the sub-leading components of the metric (in the $1/r$ expansion) can be determined in terms of $\fourshear$ and the boundary conditions at $\scrip_-$. As we show below, in $D = 6$, the free data for Einstein's equations at $\scrip$ is parametrized by $C_{ab}$ and $D_{ab}$. All the fields appearing in eq.\eqref{metric101} except for $M^{(3)}$ can be written in terms of $C_{ab}$ and $D_{ab}$ (see appendix \ref{app:graveom}). As we shall see later, $M^{(3)}$ will be identified as the analgoue of the Bondi mass (see sec.\ref{sec:bondimass}) and its $u$-derivative is determined by the constraint equation. The differences between the $D = 4$ and $D = 6$ radiative data are summarized in table \ref{table4D6D}.

%Consider the equation of motion $R_{ab} = 0$. One can study this equation at $O(r^{0})$ and $O(r^{-1})$ and obtain the behavior of $C_{ab}$ and $D_{ab}$ respectively. 
From the leading order non-trivial equation for $R_{ab}$, i.e, at $O(r^0)$ we get,
\be
\p_u C_{ab} = 0.
\label{eq:Cab}
\ee
 This implies that $C_{ab}(z)$ is $u$-independent and hence has no radiative information. In six dimensions the radiative degree comes with a $1/r^2$ fall off and hence is expected to be contained in $D_{ab}$.  The subsequent equations explicitly show that $D_{ab}(u, z)$ is unconstrained. Thus $D_{ab}$ is analogous to the shear field in four dimensions and the  corresponding News tensor $\p_u D_{ab}$ captures radiative content. %This result has been analyzed in the linearized theory \cite{Aggarwal:2018ilg} and as we will see in section \ref{sec:bondimass}, this is indeed the case even in the full theory. 

We note that as $\p_u C_{ab} = 0$, a field redefinition of $D_{ab} \to D_{ab} + {\rm \bf func}(C_{ab})$ does not change the News tensor. The freedom in defining $D_{ab}$ will play a central role in our analysis. However, as we prove in section \ref{STgen}, this freedom is fixed and {\bf func} has a specific form such that certain fall-off conditions are well defined (these fall-offs refer to the behavior of the radiative field near $\scrip_-$ and are discussed in section \ref{STgen}).

We also note that the radiative degrees of freedom parametrized by (trace-free) $D_{ab}$ are exactly equal to nine, which is the number of graviton polarizations in $D = 6$.

We finally note that in four dimensions, if we restrict ourselves to the so-called Christoudoulou-Klainermann (CK) spacetimes then the leading component of the magnetic part of the Weyl tensor vanishes at $\mathcal I^{\pm}$ \cite{Christodoulou:1993uv}. 
%This leads to the so-called CK constraints on the radiative data.
In the present case, the vanishing magnetic charge condition takes the form
\be
\p_a U_b^{(0)} - \p_b U_a^{(0)} = 0. 
\label{eq:weylmagnetic-1}
\ee
It can be checked that this condition sets the leading order term in the magnetic part of the Weyl tensor to zero\footnote{The relationship between the above constraint and the magnetic part of the Weyl tensor can be found in \cite{Hervik:2012jn}, also see \cite{Campiglia:2016jdj}.}. As $H^{1}(\mathbb S^4, \mathbf R) = 0$, $U_a^{(0)}$ is an exact form on $\mathbb S^4$ if the magnetic charge vanishes (a similar argument also holds in $\mathbb R^4$). Using \eqref{eq:Uequation}, we thus see that $C_{ab}$ is parametrized in terms of a single scalar potential\footnote{This counting can even be done in the linearized theory and was mentioned in \cite{Aggarwal:2018ilg}, but we find the reasoning below \eqref{eq:weylmagnetic-1}  to be more robust than the one in \cite{Aggarwal:2018ilg}. We would like to thank Ankit Aggarwal for discussions about this issue.}.
\be
C_{ab} = -2 \big( \partial_{a} \partial_{b} \psi \big)^{\tf}
\equiv- 2 \Big( \partial_{a} \partial_{b} \psi - \frac{1}{4} \delta_{ab} \partial^{2} \psi \Big). 
\label{eq:psidef}
\ee
Here $X^{\tf}$ denotes the tracefree part of $X$ defined as $X^{\tf}_{ab} = X_{ab} - \frac{1}{4} \delta_{ab} \tr(X)$; and $\p^2 \equiv \p^a \p_a$. Here the factor of $-2$ in the equation for the potential is chosen for convenience and will be explained in section \ref{STgen}. We note that the magnetic constraint in \eqref{eq:weylmagnetic-1} can be trivially satisfied by choosing $C_{ab} = 0$, but as was shown in the linearized case \cite{Aggarwal:2018ilg}, this condition is not preserved under supertranslations (see eq.\eqref{Cabtrans}).

%%%%%%%%%%%%%%%%%%%%%%%%%%%
\subsection{Supertranslation Generators}\label{STgen}
In this section, we analyze the symmetries which preserve the space of all $D= 6$ (and in general all even) dimensional asymptotically flat geometries\footnote{We thank Prahar Mitra for several discussions regarding this.}. If we restrict the analysis to linearized gravity around flat spacetime, then it has been shown in \cite{Kapec:2015vwa, Aggarwal:2018ilg} that the asymptotic symmetries include infinite dimensional supertranslations. The complete group of such symmetries is a Cartesian product of BMS group on $\scrip$, $\scrim$. The diagonal subgroup obtained by the anti-podal identification of supertranslation generators at $\scrip$ and $\scrim$, is the symmetry group of tree-level S-matrix with massless external particles (which are not gravitons) such that the corresponding Ward identity is equivalent to the Weinberg soft graviton theorem. 

Although this is a promising data point for showing existence of an IR triangle in higher (even) dimensions, there is a caveat which has not been analysed in the literature so far\footnote{Henceforth, we shall always mean higher even dimensions, whenever we refer to any results in higher dimensions.}. As seen in the literature \cite{Kapec:2015vwa, Aggarwal:2018ilg} and also discussed previously, the action of supertranslation even in linearized gravity generates an infinity of flat spacetimes parametrized by $C_{ab} \neq 0$. Hence in order to show that supertranslations at $\mathcal I^{\pm}$ preserve the space of linearized asymptotically flat geometries, we need to include linearisation around flat space-times for which $C_{ab} \neq 0$. To the best of our understanding, analysis of supertranslation symmetries and the subsequent conservation laws  have not been studied previously and we fill this gap in our paper.

 We start by reviewing the derivation of BMS asymptotic symmetries as given in \cite{Kapec:2015vwa, Aggarwal:2018ilg}. They are generated by vector fields which are asymptotically Killing. I.e, $\xi^\a$ is an asymptotic symmetry if, 
\begin{eqn}
 \mathcal{L}_{\xi} g_{rr} = 0, \qquad 
 \mathcal{L}_{\xi} g_{ra} = 0, \qquad 
 g^{ab}\mathcal{L}_{\xi} g_{ab} = 0. 
\end{eqn}
We note that this definition of asymptotic symmetry is agnostic to the dimension of spacetime. These equations tell us that the gauge conditions should be preserved under a diffeomorphism. They can be solved exactly and the solutions for $\xi^\mu$ is given as \cite{Kapec:2015vwa}, 
\begin{eqn}
 \xi^u = f, \qquad \xi^a = - \p_b f \int_{r}^\infty e^{2\b} g^{ab} dr' , \qquad \xi^r = \frac{r}{4}\Big( U^a \p_a f - \p_a \xi^a \Big),
\end{eqn}
which can be summarized as,
\be\label{Xi}
\xi^\mu = f \p_u + \frac{r}{4} \Big( U^a \p_a f - \p_a \xi^a \Big) \p_r - \p_b f \int_r^{\infty} e^{2\beta} g^{ab} dr' \p_a .
\ee
These are the well known BMS vector fields which include supertranslations parametrized by an arbitrary function $f(z^a)$\footnote{If we relax the boundary conditions and demand that the asymptotic symmetries are generated by vector fields which are volume preserving as opposed to asymptotically Killing, then we will be led to the generalisation of the so-called Generalised BMS group which is the semi-direct product ${\rm ST}\, \otimes \,  \textrm{Diff}(S^{4})$. We expect the conservation laws associated to $\textrm{Diff}(S^{4})$ to be equivalent to sub-leading soft graviton theorem. However the detailed analysis of this question is outside the scope of this paper.}. We now derive the action of the supertranslations on the radiative data at $\scrip$. By evaluating $\mathcal L_\xi g_{ab}$ (which we shall often denote as $\delta_f g_{ab}$ when we are referring to supertranslations) to $O(r^{-1})$ we can see that, 
\be
\lim_{r \to \infty} r \mathcal L_\xi g_{ab} = \delta_f C_{ab} = - 2 \Big(\p_a \p_b f - \frac{1}{4} \delta_{ab}\p^2 f \Big). 
\label{Cabtrans}
\ee
where $\xi$ is the ST vector field defined in \eqref{Xi}. It can be verified immediately that the transformation of the potential $\psi$ in \eqref{eq:psidef} is as follows\footnote{This explains our reason for introducing -2 in the definition of the scalar potential in \eqref{eq:psidef}. Without which we would have to rescale $\psi$ in an appropriate manner.}
\be
\delta_f \psi = f
\ee 
We can now compute the action of ST on $D_{ab}$,
\begin{eqn}\label{Dabvariation}
 \lim_{r \to \infty} r^2 \mathcal L_\xi g_{ab} = \delta_f D_{ab} &= f \p_u D_{ab} + \frac{1}{4} \delta_{ab} \left[  -\frac{4}{3}  \p_c C^{cd} \p_d f - C^{cd} \p_c \p_d f \right] \\
 &\quad + \frac{1}{4} C_{ab} \p^2 f - \p_c C_{ab} \p^c f  - \frac{1}{2} \big( C_{bc} \p_a \p^c f + C_{ac} \p_b \p^c f \big) \\
 &\qquad + \frac{1}{2} \big[ \p_a C_{bc} \p^c f + \p_b C_{ac} \p^c f \big]+ \frac{1}{6} \big[ \p^c C_{bc} \p_a f + \p^c C_{ac} \p_b f \big]. 
\end{eqn}
This result is completely general and is valid in the non-linear theory. From the perspective of linearized gravity, it is an extension of the result in \cite{Aggarwal:2018ilg} when the background spacetime has non-zero $C_{ab}$\footnote{From the perspective of soft theorems, our analysis can be understood as a derivation of the leading order multi-soft graviton theorem from supertranslation Ward identity in linearized gravity.}.

We thus note that the action of ST on $D_{ab}(u, z)$ (Graviton mode) generates a constant ($u$-independent) term even in {\it linearized gravity around a flat metric with non-zero $C_{ab}$\footnote{An analysis about a flat metric with non-zero $C_{ab}$ means that the background is $r^2 \delta_{ab} + r C_{ab}$ instead of just $r^2 \delta_{ab}$}}. However, both the saddle point analysis in appendix \ref{app:saddle} and the finiteness of symplectic structure (see section \ref{sec:symplecticform}) at $\scrip$ impose stronger fall off conditions on the Graviton mode. The finiteness of the symplectic structure implies that as $|u| \to \infty$
\be
\lim_{|u| \to \pm \infty} {\rm Graviton}(u, z) = O\Big( \frac{1}{|u|^{1 + 0_+}} \Big)
\ee
 This fall-off condition appears to be in tension with \eqref{Dabvariation}. We now show that it is possible to redefine $D_{ab}$ to $\tilde D_{ab}$ such that (1) the corresponding News tensor $\p_u D_{ab}$ remains unchanged and (2) action of ST preserves the fall-off condition of $\tilde D_{ab}$. Let us define $\tilde D_{ab}$ as,
\be
\tilde D_{ab} = D_{ab}  - \frac{1}{4} C_a^c C_{bc} - \frac{1}{16} \delta_{ab} C^{cd} C_{cd}.
\label{Dtildedefn}
\ee
Under ST it transforms as, 
\be\label{Dtiltrans}
\delta_f \tilde D_{ab} = f \partial_u \tilde D_{ab}. 
\ee
It can also be verified that $\tilde D_{ab}$ is trace-free, i.e, $\tilde D_a^a =0$. Implications of this field redefinition in linearized gravity are now clear.  When combined with the results of \cite{Kapec:2015vwa}, we see that the Gravitational scattering in the background of any flat spacetime has ST symmetries whose associated conservation laws are equivalent to classical and quantum soft graviton theorem \cite{Laddha:2019yaj}. The complete implications of this field redefinition in non-linear theory will emerge in the following sections. For now, we have the following statement. In the non-linear theory, supertranslations map an asymptotically flat spacetime to a distinct asymptotically flat spacetime via the following action, 
\begin{subequations}
\begin{align}
 \delta_f C_{ab} &= - 2 \Bigl( \partial_{a} \partial_b f- \frac{1}{4} \delta_{ab} \partial^2 f \Bigr), \\
 \delta_f \tilde D_{ab} &= f \partial_u \tilde D_{ab}.
 \end{align}
\end{subequations}

Having described the generators of ST, we move onto computing the charge corresponding to this symmetry in the next section.

%%%%%%%%%%%%%%%%%%%%%%%%%%%
\section{Supertranslation Charge} \label{sec:STcharge}
%%%%%%%%%%%%%%%%%%%%%%%%%%%
In this section, we evaluate the Noether charge corresponding to supertranslations. We first compute the symplectic form of the theory at $\scrip$ and then compute the Noether charge using that. Finally, we also compute the Bondi mass and explain how we obtain the same Noether charge as computed before.

%%%%%%%%%%%%%%%%%%%%%%%%%%%
%%%%%%%%%%%%%%%%%%%%%%%%%%%
\subsection{Symplectic Form}\label{sec:symplecticform}
 In order to compute the Noether charge we use the  covariant phase formalism and first analyze the symplectic form of the theory\footnote{For a detailed discussion of this formalism, we refer the reader to  \cite{Compere:2018aar, Crnkovic:1986ex, He:2020ifr}. This was first analyzed near null infinity in \cite{Ashtekar:1981bq, Ashtekar:1981sf}.}. This will also help us in the identification of the canonically conjugate variables of the theory.
 
 The Symplectic form can be constructed on any Cauchy slice but since our interest lies in the theory at future null infinity, we will perform our analysis on $\scrip\cup i^+$. The advantage  of doing this is that most interaction terms die off near that region. Since we are working with massless fields, we can neglect the contribution from $i^+$ and just work by defining the data on the null slice $\scrip$. 

We have already established in the previous sections that $C_{ab}$ and $\tilde D_{ab}$ constitute the free data and it would therefore make sense to construct the symplectic form with these two variables. 

The symplectic form in higher dimensions at $\scrip$ has been evaluated in the linearized limit about $C_{ab} = 0$ in \cite{Aggarwal:2018ilg}. We  follow the same strategy and evaluate this in the non-linear theory for the Einstein-Hilbert action at $\scrip$. In order to do this we first construct the symplectic current and later, integrate over the current to obtain the symplectic form. Following \cite{Crnkovic:1986ex}, the symplectic current for the Einstein-Hilbert action is given as\footnote{A detailed derivation of this expression can be found in section 4.2 of \cite{Grant:2005qc}. There is an improved version of this result given in \cite{Harlow:2019yfa} which takes care of boundary terms on a spatial slice.}, 
\be
J^\alpha =\frac{1}{2}\delta \Gamma^\a_{\mu\nu}\wedge \big[ \delta g^{\mu\nu} + \frac{1}{2} g^{\mu\nu} \delta \log g \big] - \frac{1}{2} \delta \Gamma_{\mu\nu}^{\nu} \wedge \big[ \delta g^{\a \mu} + \frac{1}{2} g^{\a \mu} \delta \log g \big] 
\label{eq:gravform-1}
\ee
Here $\delta$ represents the exterior derivative on phase space and $\wedge$ denotes the wedge product. The basic idea of computing this symplectic form at $\scrip$ follows a similar strategy as QED (see appendix \ref{app:QEDform}) or the linearized gravitational case, which is already done in \cite{Aggarwal:2018ilg}. We explain the technicalities of evaluating this in the non-linear gravitational case in full detail in appendix \ref{app:gravform} and summarize the conceptual points here.

The symplectic form is defined as an integral over the symplectic current, 
\be
\Omega \equiv \int d\Sigma_\alpha J^\alpha
\ee
with $d\Sigma_\alpha$ denoting the measure of integration on the chosen Cauchy slice. For us, this is evaluated on $\scrip$. In order to reach $\scrip$, we first choose a constant-$t$ cauchy slice and then take the limit $t \to \infty$ by holding $u =$ constant. Note that in our notations $t = u + r$. Since we start with a constant-$t$ Cauchy slice, the component of the current that contributes to the symplectic form is $J^t = J^u + J^r$,
\be\label{formcurrent}
\Omega^t = \lim_{r \to \infty} \intI r^4 du d^4z \ J^t = \lim_{r \to \infty} \intI r^4 du d^4z \ (J^u + J^r) .
\ee
It is simpler to evaluate $J^u$ first and that is given as (refer to \eqref{Jugrav} for more details), 
\be
-J^u = \frac{1}{4r^4} \delta C^{ab} \wedge\delta ( D_{ab} - C^a_c C^{bc} )
\label{eq:Jugrav}
\ee
For $J^r$, we get a leading order term at $O(r^{-3})$ which we call $J^r_{div}$. This term can potentially diverge since the measure of the integral in \eqref{formcurrent} at $\scrip$ is $r^4$ and we are working in the large-$r$ limit. However, as shown in \eqref{app:Jrdiv}, it ends up giving a finite contribution to the symplectic form\footnote{We emphasize that this apparent large-$r$ divergence is also seen in QED \eqref{Judivem} and in the linearized limit as well \cite{Aggarwal:2018ilg}. As shown here, this divergence disappears upon appropriately choosing the boundary conditions.}. Thus the contribution from $J^r$ is given as,
\be
J^r = J^r_{div} + J^r_{fin}
\ee
with 
\begin{subequations}
\be\label{Jdiv-1}
-2J^r_{div} =  - \frac{1}{2r^4} \p_u \Big[ (t - u) \delta C^{ab} \wedge \delta  D_{ab} \Big] - \frac{1}{2r^4} \delta C^{ab} \wedge \delta  D_{ab} ~,
\ee
\begin{eqn}
 - 2J^r_{fin}&=-\frac{1}{2} \delta D^{ab} \wedge \delta \p_u D_{ab}  - \frac{1}{2} \delta C^{ab} \wedge \delta \p_u E_{ab}\\
  &\quad-\frac{1}{2} \delta C^{ab} \wedge \delta \bigg[ 2 \p_{a} U_{b}^{(1)}  - U^{(0)c} \big( 2\p_a C_{bc}  - \p_c C_{ab} \big) - \frac{1}{3} \p_b\big[ U_a^{(1)} + C_a^c U_c^{(0)} \big] \\
  &\qquad \qquad \qquad\quad  + \frac{1}{3} \p_m \p_a (D_b^m - C_{c}^m C^{c}_b) \bigg] .
  \label{Jrfin1}
\end{eqn}
\end{subequations}
The equations can be simplified further by expressing it in terms of the free data $C_{ab}$ and $\tilde D_{ab}$ (see appendix \ref{app:gravform} for details). We  then combine $J^r$ and $J^u$ to obtain $J^t$ which we integrate to get the symplectic form \eqref{formcurrent}. In order to ensure a finite contribution from the first term in \eqref{Jdiv-1} to the symplectic form (in the large$-r$ limit), we see that the fall off condition on $\tilde D_{ab}$ with large$-u$ should be\footnote{A similar analysis also motivates the fall off in the linearized theory, as shown in \cite{Aggarwal:2018ilg}, and can also be seen in QED \eqref{A1ifalloff}.} 
\be\label{Dtilfalloff}
\lim_{u \to \pm \infty}\tilde D_{ab} = O\Big(\frac{1}{|u|^{1 + 0_+}}\Big).
\ee
For further details we refer the reader to the appendix \ref{app:Jrdiv}.  We also note that the same fall-off is expected from the saddle point analysis, as explained in appendix \ref{app:saddle}.

Leaving the details to appendix \ref{app:Jtfull}, we state the final result for the the symplectic form here. We find that the symplectic form $\Omega^t$ can be uniquely spit into two different terms
\begin{enumerate}
 \item {\bf Finite and Integrable piece:} This is the term that leads to the correct Noether charge and also defines the radiative phase space at $\scrip$, 
 \be
 \Omega^{rad} =  \intI du d^4 z J^t_{I} = - \frac{1}{2}\intI d^4 z du \Big[- \frac{1}{2} \delta \tilde D_{ab} \wedge \delta\p_u \tilde D^{ab} + \frac{1}{2} \delta C^{ab} \wedge \delta \p_u E_{ab}+ \frac{1}{9} \delta C^{ab} \wedge \delta  \p_a \p^c \tilde D_{bc} \Big]. 
\label{omegatfin}
 \ee
 It contains the radiative degree of freedom $\tilde D_{ab}$ and defines its Poisson brackets near $\scrip$. We thus obtain a generalization of the Ashtekar-Streubel symplectic structure at $\scrip$ \cite{Ashtekar:1981bq} in the higher dimensional non-linear theory. We also note that the expression above can be obtained from \cite{Aggarwal:2018ilg} by replacing $D_{ab}$ with $\tilde D_{ab}$, which is the radiative data in the non-linear theory. Therefore upon using the EOM for $E_{ab}$ (see eq.\eqref{eq:Eabeom}), the canonically conjugate variable of $\tilde D_{ab}$ can be identified using this structure. As shown in section \ref{sec:sympcharge}, we obtain the Noether charge of supertranslation using $\Omega^{rad}$. Later (in section \ref{sec:bondimass}), we also compare it with the expression obtained from the Bondi mass. The divergent piece as discussed below does not affect the charge as it cannot be expressed as a total variation. 
 
 \item  {\bf Divergent and Non-Integrable piece:} The other piece of $\Omega^t$ is given as
 \begin{eqn}
\Omega^{div}&=   \frac{1}{2}\intI du d^4 z J^t_{NI}\\
&= - \intI du d^4 z \ \delta C^{ab} \wedge \delta \Big[-\frac{1}{2} C_a^c C_{bc}-\frac{1}{9} \p_a \Big(  C^{c d} \p_c C_{b d} + \frac{4}{3} C_b^c \p^d C_{c d} - \frac{1}{16} C^{cd} \p_b C_{cd}  \Big) \\
 &\hspace{4.5cm} + \frac{1}{3} \p^c C_{cd} \p_a C^d_b - \frac{1}{6}\p^c C_{ab} \p^d C_{cd} \Big].
 \label{omegatNT}
\end{eqn}
Note that this contains no radiative degree of freedom $\tilde D_{ab}$, and as $\p_u C_{ab} = 0$, the integrand is independent of $u$. Therefore, this gives a divergent term when integrated with $u$.  However, this does not make the symplectic structure ambiguous at $\scrip$. The ambiguity is settled by noting that this term will not contribute to the Noether charge, as it does not lead to a total variation. As seen from the computation in \cite{Aggarwal:2018ilg}, such a term does not arise while computing the symplectic form in the linearized theory about the $C_{ab} = 0$ background.  It will be interesting to see the implication of such a term and we leave that for a future work.

\end{enumerate}

%%%%%%%%%%%%%%%%%%%%%%%%%%%
\subsection{Noether Charge} \label{sec:sympcharge}
The Noether charge\footnote{Charges corresponding to gauge symmetries require a study of the generalized Noether theorem, which is reviewed in \cite{Compere:2018aar}. } can be computed using 
\be
\delta Q_f \equiv  \Omega^{rad}(\delta_f,  \delta) 
\label{sympcharge}
\ee
Here $\delta_f$ represents the ST variations\footnote{For a computation of the charge in QED, refer to appendix \ref{app:QEDcharge} and also \cite{Freidel:2019ohg, Henneaux_2019}.}. The following calculation is almost similar to that of \cite{Aggarwal:2018ilg} with $D_{ab}$ replaced by $\tilde D_{ab}$.  However in addition to the soft charge we also obtain the hard charge as shown below. 

We first work out the soft charge (the term linear in $\tilde D$). Using the formula \eqref{sympcharge}, the soft charge obtained from the symplectic form becomes,
\be\label{gravsympcharge}
Q_f^{soft} = \frac{1}{12} \intI d^4 z du \ f \p^2 \p^{ab} \tilde D_{ab} +\intI d^4 z du \ f ( \p^{ab} - \frac{1}{4} \delta^{ab} \p^2) \p_u E_{ab}
\ee
where $\p^{abc\cdots} \equiv \p^a \p^b \p^c \cdots$. From the equation of motion of $E_{ab}$  (see eq.\eqref{eq:Eabeom}) it is easy to see that the term containing $\p_uE_{ab}$ does not contribute,
\be\label{papbEab}
\big(\p^{ab} - \frac{1}{4}\delta^{ab} \p^2 \Big) \p_u E_{ab}=  \p^{ab}\big(C^c_{(a} \p_u \tilde D_{b) c} \big) - \frac{1}{4} \p^2 \big(C^{ca} \p_u \tilde D_{ca}\big)
\ee
 Using this and the fall condition \eqref{Dtilfalloff}, it is clear that such a term does not contribute to the charge, as upon integrating \eqref{papbEab} w.r.t $u$, we get zero. The other  term in \eqref{gravsympcharge} is similar to the expression obtained in \cite{Aggarwal:2018ilg}, with the identification $D_{ab} \to \tilde D_{ab}$. The only other difference is that we are working with the flat sphere $\delta_{ab}$, and hence we do not see the contribution from the curvature term of $\mathbb S^4$ that arose in \cite{Aggarwal:2018ilg}. Therefore, the soft charge reduces to
\be
Q_f^{soft} = \frac{1}{12} \intI du d^4 z  \ f(z) \p^2 \p^{ab} \tilde D_{ab}.
\ee
The total charge is obtained by adding this to the contribution from the matter and hard gravitons (also known as the {\it hard charge}), which is given as, 
\be\label{totalcharge}
Q_f = Q_f^{soft}+ Q_f^{hard} = Q_f^{soft} +\intI du d^4 z \ f(z) \Big(T_{uu}^{M(4)} + \frac{1}{4} N_{ab} N^{ab} \Big)
\ee
where we explicitly state the contribution from the matter and the graviton part. The graviton hard charge is obtained from the following term in the symplectic form \eqref{omegatfin} and by using the variation \eqref{Dtiltrans} 
\bes
\int d^4 z du \ \delta \tilde D_{ab} \wedge \delta \p_u \tilde D^{ab}.
\ees

 We can also obtain the same expression for the charge by using the Bondi mass, as shown in the next subsection (see section \ref{sec:bondimass}). We see that our final result \eqref{totalcharge} is a neat generalization of \cite{Kapec:2015vwa, Aggarwal:2018ilg} to the non-linear theory with $D_{ab}$ replaced by $\tilde D_{ab}$.

In the following section we demonstrate how we get the total charge \eqref{totalcharge} using the Bondi mass.

%%%%%%%%%%%%%%%%%%%%%%%%%%%
%%%%%%%%%%%%%%%%%%%%%%%%%%%
\subsection{Bondi Mass}\label{sec:bondimass}
We first give a brief idea of the Bondi mass in the six dimensional non-linear theory, explain the notations, and then demonstrate how it generates the expression for the supertranslation charge \eqref{totalcharge}.

The {\it Bondi mass aspect}, $M^{(3)}(u, z)$ is denoted by the notation $m_B(u, z)$. At $\scrip$ this gives us a definition of the angular density of energy. For Kerr-like spacetimes the Bondi mass aspect is proportional to the mass of the object in the bulk. We obtain the {\it total Bondi mass} by integrating the Bondi mass aspect over the Celestial sphere (or plane). And further, the total Bondi mass in the limit $u \to -\infty$ gives us the ADM mass, which is identified as the Hamiltonian of the theory. 

We can study the evolution of the Bondi mass aspect along $\scrip$ using the $uu-$component of the Einstein equation (we use the same conventions as that of \cite{Aggarwal:2018ilg}),
\begin{eqn}
 &-2 \p_u m_B - \frac{1}{4} \p_u D^a_b \p_u D^b_a - \p_u \p_a U^{(2) a}  + \p_a \big(  C^{ab} \p_u U_b^{(1)} \big)- 3 U^{(0)a} \p_u U_a^{(1)} \\
 &=  T_{uu}^{M(4)} + \frac{1}{2}\p^2 (U^{(0)2} + M^{(2)} )
 \label{mBevolution}
\end{eqn}
Here we have also included a matter source term as denoted by $T_{uu}^{M(4)}$. Using the EOM for $M^{(2)}$ and $U_a$'s (as given in appendix \ref{app:graveom}), we can express $\p_u m_B$ entirely in terms of the free data $C_{ab}$ and $\tilde D_{ab}$. Further, it is simple to see that in the linearized limit, equation \eqref{mBevolution} reduces to the results in \cite{Kapec:2015vwa, Aggarwal:2018ilg}. In order to get the total Bondi mass, we need to integrate $\p_u m_B$ over  $z^a$. Since we are not considering massive particles in our system, there is no non-trivial information at $u \to +\infty$ and hence $m_B(+\infty, z) = 0$. Using this information and eq.\eqref{mBevolution}, the total Bondi mass is given as,
\be
\int d^4 z\ m_B(u, z) =  \frac{1}{2}\int_{u}^\infty du d^4 z \  T_{uu}^{M(4)} + \frac{1}{4} \p_u \tilde D^{ab} \p_u \tilde D_{ab}. 
\ee
The ADM mass can be obtained by taking the $u \to -\infty$ limit in the equation above, 
\be
\lim_{u \to -\infty}\int d^4 z\ m_B(u, z) = \frac{1}{2}\int_{-\infty}^\infty du d^4 z \ T_{uu}^{M(4)} + \frac{1}{4} \p_u \tilde D^{ab} \p_u \tilde D_{ab}. 
\label{adm}
\ee

Having explained the basic properties of $m_B(u, z)$, we now describe its relevance as a candidate for the supertranslation charge. This is motivated as a valid candidate through the results in the linearized theory \cite{Kapec:2015vwa, Aggarwal:2018ilg}, and is given as
\be
\mathcal Q_f \equiv \lim_{u \to -\infty} 2\int f(z) m_B(u, z) d^4 z 
\label{Qst}
\ee
where we have an extra factor of 2 since we set $8\pi G_N = 1$\footnote{The expression without working with $8 \pi G_N = 1$ is given as \cite{Kapec:2015vwa}\bes  \mathcal Q_f \equiv \lim_{u \to -\infty} \frac{1}{4\pi G_N}\int f(z) m_B(u, z) d^4 z 
\ees} and $f(z)$ is the function parametrizing supertranslations (see section \ref{sec:STshigher}). We can recover the ADM mass \eqref{adm} by setting $f(z) = {\rm constant}$.%\footnote{On the regular sphere $\mathbb S^4$, the function $f(z)$ is proportional to $Y_0(z)$ where  $Y_l$ is the spherical harmonic in 4-dimensions}. 

Substituting the constraint equation \eqref{mBevolution} in \eqref{Qst} and after integrating by parts, we obtain $\mathcal Q_f$ in terms of the radiative data, which is expressed as a summation of two separate pieces. Comparing with equation \eqref{totalcharge}, these are identified as the {\it soft} and {\it hard} piece
\be\label{noethercharge}
\mathcal Q_f = \mathcal Q^{soft}_f +\mathcal  Q^{hard}_f
\ee
where, 
\begin{subequations}
 \be
 \mathcal Q_f^{soft} = \frac{1}{12} \intI d^4 z du \ f(z)  \p^2 \p^{ab} \tilde D_{ab}
 \label{eq:softcharge}
 \ee
 \be
 \mathcal Q_f^{hard} = \intI d^4 z du \ f(z)  T_{uu}^{(4)}.
 \label{eq:hardcharge}
 \ee
\end{subequations}
 The hard charge contains the stress tensor of matter and hard gravitons, which is given as
 \be\label{stresstensor}
  T_{uu}^{(4)} =  T_{uu}^{M(4)} + \frac{1}{4} \p_u \tilde D^{ab} \p_u \tilde D_{ab}.
 \ee
 Therefore we see that the expression derived in \eqref{noethercharge} and the one derived using the symplectic form (see eq.\eqref{totalcharge}) are the same. As noted before, the charge derived in this section, generalizes the results in \cite{Kapec:2015vwa, Aggarwal:2018ilg} to the non-linear theory (and also to the linearized theory about a $C_{ab} \neq 0$ background). From the discussion above, we see that it is possible to formally obtain the result for the charge in the non-linear theory by using the results in \cite{Kapec:2015vwa, Aggarwal:2018ilg} and replacing $D_{ab}$ with $\tilde D_{ab}$. This is another way of deriving the redefined Graviton mode $\tilde D_{ab}$.  Therefore, combining with the results of \cite{Kapec:2015vwa}, we expect to get a similar structure for the Ward identity and correspondingly, the leading soft-theorem in the non-linear theory. We elaborate on this point and discuss some more implications of the Ward identity in section \ref{sec:discussion}.

We note that in the linearized theory about $C_{ab}= 0$, the total charge $Q_f$ has also been derived from the electric part of Weyl tensor in \cite{Aggarwal:2018ilg} and it will be interesting to compute the same in the non-linear theory and verify that we get the same result as \eqref{noethercharge}, which we leave for a future work.

This completes our analysis of the Supertranslation charge and it also sheds light on the phase space of the theory. Along with the description of asymptotic symmetries and soft theorems, there is another important ingredient alluding to the infrared properties of the theory. This is known as the {\it memory effect} and together with the soft theorems and asymptotic symmetries, it completes the IR triangle. The triangle represents a set of mathematical operations which connects the three corners. We now move onto the analysis of the memory effect and describe how it is generalized from the linearized results and also, how it is related with the discussions in the sections above.

%%%%%%%%%%%%%%%%%%%%%%%%%%%
\section{Memory and IR Triangle}\label{sec:memory}
%%%%%%%%%%%%%%%%%%%%%%%%%%%
In \cite{Pate:2017fgt}, the authors have defined the linearized memory in (even) higher dimensions. In 6-dimensions,  it is proportional to $D_{ab}(u, z)$. The existence of memory follows directly from the equation of motion and has been proved in \cite{Laddha:2018rle}. The fact that memory only depends on the scattering data and not on details of the interaction is the statement of classical soft graviton theorem. However the results of \cite{Laddha:2018rle} are valid for linear as well as non-linear memory (also known as null memory \cite{Bieri_2014}). In this section, we will show that the non-linear memory can be obtained directly from $\tilde D_{ab}$ (defined in \eqref{Dtildedefn}) such that in the linearized theory around Minkowski vacuum, it reduces to the linear memory derived in \cite{Pate:2017fgt}.

A definition of the memory effect convenient for our purpose is the following. Memory effect is a measure of how the distance between two detectors near $\scrip$ changes upon the passage of gravitational radiation. These detectors move in a time like trajectory with the tangent vector $k = \p_u$. The relative separation of the detectors is usually computed using the Geodesic deviation equation, 
\be
{d^2 s^a \over d u^2} = R^a_{\ u u b} s^b
\label{geodesicdeviation}
\ee
where $s^a$ is the relative transverse displacement between the two detectors and $R^a_{\ u u b}$ is the Riemann curvature. This equation gives us $s^a$ as a function of $u$ and we will solve this equation perturbatively in $G_N$, i.e, assume that the $s^b$ on the RHS of \eqref{geodesicdeviation} is at a fixed value $u$ and then study the separation by recursively solving \eqref{geodesicdeviation}. For our purpose, the leading order  solution to this recursion equation will suffice. We note that the detectors can make a measurement between retarded times $u_i$ and $u_f$ and therefore, the memory is defined as $s^a(u_f) - s^a(u_i) \equiv \Delta s^a$. This represents the change in separation between the detectors in the retarded time interval $u_f - u_i \equiv \Delta u$. To account for gravitational radiation at late and early times, memory is usually defined in the limit $u_i \to -\infty$ and $u_f \to +\infty$. 

In order to study $\Delta s^a$ via \eqref{geodesicdeviation}, we first compute the Riemann tensor $R^a_{\ u u b}$. Note that the indices of $R^a_{\ u u b}$ are raised and lowered using $g_{\mu\nu}$ but the indices of $C_{ab}$, $\tilde D_{ab}$ and $E_{ab}$ are raised and lowered using $\delta_{ab}$. In the large-$r$ limit $R^a_{\ u u b}$ is given as, 
\be
 R^a_{\ u u b} = \frac{1}{2r^2} \p_u^2 \tilde D^a_b + \frac{1}{2r^3} \Big( \p_u^2 E^a_b - \frac{2}{3} \p_u  \p_c^{(a} \tilde D^c_{b)} - C^{ac} \p_u^2 \tilde D_{bc} + \frac{1}{6} \delta^a_b \p_u \p^{cd} \tilde D_{cd}\Big) + O(r^{-4}).
\ee
This can be expressed in terms of the radiative data by using the EOM for $E_{ab}$, 
\be
\p_u E_{ab}= C_{(a}^c \p_u \tilde D_{b)c} + \frac{2}{3} \p_{(b}^c \tilde D_{a) c} - \frac{1}{2} \p^2 \tilde D_{ab} - \frac{1}{6} \delta_{ab} \p^{cd} \tilde D_{cd}. 
\label{eq:Eabeom}
\ee
Using this we find, 
\be
R^a_{\ u u b} = \frac{1}{2r^2} \p_u^2 \tilde D^a_b - \frac{1}{4r^3} \Big( \p^2 \p_u \tilde D^a_b +  C^{ca} \p_u^2 \tilde D_{b c} - C^{c}_b \p_u^2 \tilde D^{a}_{c} \Big)+ O(r^{-4})
\ee
Integrating this twice with $u$ gives us the memory $\Delta s^a$ (see \eqref{geodesicdeviation}) to leading order in $r$, 
\be \label{nonlinmem}
\Delta s^a = - \frac{1}{4 r^3} \int_{-\infty}^{\infty} du \ \p^2 \tilde D^a_{\ b} \ s^b_i
\ee
where we have used the boundary condition $\tilde D_{ab}(u \to \pm\infty) \to 0$ (see \eqref{Dtilfalloff}) and $s^b_i \equiv s^b(u_i)$ with $u_i \to -\infty$. The boundary condition sets the term at $O(1/r^2)$ and the non-linear term at $O(1/r^3)$ to zero.

This generalizes the definition of the linear memory to the non-linear theory in six dimensions. Thus we get a contribution of the six dimensional memory at the Coulombic order, in contrast with the four dimensional case (see table \ref{table4D6D} for a summary the differences between the important physical quantities in four and six dimensions).  In appendix \ref{app:memory}, we show how one can recover a similar form of the answer from the computation in \cite{Pate:2017fgt} in the linearized regime.

From \eqref{nonlinmem} we see that the memory in the non-linear theory is measured via an integration over the radiative degree of freedom $\tilde D_{ab}$. This is the precisely the same quantity which appears in the soft charge (see \eqref{eq:softcharge}) and eventually, in the soft theorems as well (we expand upon how one gets to the soft theorems using the conservation laws in section \ref{sec:discussion}). Upon comparing with the results in the linearized theory \cite{Kapec:2015vwa, Aggarwal:2018ilg, Pate:2017fgt}, we see how our results are generalized to the non-linear theory via the replacement of $D_{ab} \to \tilde D_{ab}$. This establishes the connection between the three corners of the IR in the full non-linear theory. Therefore, we see how the IR triangle is generalized to the full non-linear theory of gravity in higher-dimensions.

%%%%%%%%%%%%%%%%%%%%%%%%%%%
\section{Discussion and summary}\label{sec:discussion}
%%%%%%%%%%%%%%%%%%%%%%%%%%%
We summarize our main results and state some important implications for the quantum theory in this section. 
%%%%%%%%%%%%%%%%%%%%%%%%%%%
\subsection{Possible Implications for the Quantum theory}\label{sec:quantumdisc}
As discussed in section \ref{sec:STshigher}, the free data of the theory is constituted by $C_{ab}$ and $\tilde D_{ab}$, where the mode $C_{ab}$ is $u$-independent and $\tilde D_{ab}$ is identified as the shear. The $u$-independent mode labels the vacuum of the theory. It has been well known \cite{Strominger:2017zoo} that for a theory having an infinite dimensional BMS symmetry, there are multiple vacuua states possible. As given in equation \eqref{Cabtrans}, the generators of supertranslations modify the value of $C_{ab}$ so it is identified as the Goldstone mode and since the theory has infinite dimensional BMS symmetry it has multiple vacuua. 

A table summarizing the comparison with 4D is given in table \ref{table4D6D}. 
\begin{table}[h]
 \centering
 \begin{tabular}{{ | m{1.65cm} | m{1.5cm}| m{1.5cm} | }}
\hline
 {\bf Quantity} & {\bf 4D} & {\bf 6D}\\
 \hline
 Goldstone Mode& $\int C_{ab}^{(4D)} du $ & $C_{ab}$ \\
 \hline
 Radiative Mode & $C_{ab}^{(4D)}$ & $\tilde D_{ab} $ \\
 \hline 
 Memory & $C_{ab}^{(4D)} $ & $\tilde D_{ab} $\\
 \hline
 Soft Mode & $\int C_{ab}^{(4D)} du$ & $\int \tilde D_{ab} du$ \\ 
 \hline
\end{tabular}
\caption{Here we summarize the importance of the components in the $r$-expansion of $g_{ab}$ and show how they contrast with the 4D counterpart. Note that the quantities given are only up to a proportionality, and focuses on the $u$-dependence. }
\label{table4D6D}
\end{table}

We quote our first main result in this language: The true graviton degree of freedom gets redefined upon working about a specific vacuum (labeled by $C_{ab}$). From the structure of \eqref{Dtildedefn} we see that it is specifically the zero-mode of $D_{ab}$ which gets redefined.

We now describe an important application of our results in the context of the S-matrix\footnote{We would end up with a similar implication even for the QED S-matrix in higher dimensions.}. As described above, there exists multiple soft vacuua in flat space and thus the initial and the final states in a generic S-matrix can be built on different soft vacuua. Therefore, there is a possibility of a vacuum-vacuum transition in a generic scattering process. There is a detailed calculation demonstrating this effect in a four dimensional scattering process \cite{Choi:2017ylo}. The main reason behind this transition is the {\it Ward Identity}, which, for a scattering process would imply that the total supertranslation charge (see eq.\eqref{noethercharge}) is conserved during scattering, 
\begin{eqn}\label{ward}
&\langle out|[\hat Q_f, S]|in \rangle = \langle out|[\hat Q_f^{hard} + \hat Q_f^{soft}, S]|in \rangle = 0\\
&\implies  (Q_{f, +}^{soft} - Q_{f, -}^{soft})\braket{out|S|in} =- (Q_{f, +}^{hard} - Q_{f, -}^{hard})\braket{out|S|in}
\end{eqn}
where we use the notation $\hat Q |out\rangle = Q_+ |out\rangle$ and $\hat Q |in\rangle = Q_- |in \rangle$. We can consider an $|in\rangle$ state built on an eigenstate of $\hat Q_{f}^{soft}$ with eigen charge $Q_{f, -}^{soft}$. 
For any general scattering with a non-zero S-matrix, $\braket{out|S|in} \neq 0$ we have $Q_{f, +}^{hard} \neq Q_{f, -}^{hard}$. Therefore for a scattering process which conserves the supertranslation charge (satisfies \eqref{ward}), we must have $Q_{f, +}^{soft} \neq Q_{f, -}^{soft}$. This indicates that there is a vacuum to vacuum transition in any general scattering which conserves the supertranslation charge.

 It is well known that the S-matrix in a theory of gravity in 4-dimensions suffers from IR-divergences \cite{Weinberg:1965nx}. However the physical S-matrix (which is free of IR divergences) can be obtained by dressing the original S-matrix using the KF (Kulish-Faddeev) prescription \cite{Kulish:1970ut}. Therefore, the IR finite S-matrix in 4-dimensions is given by the KF dressed S-matrix. This leaves us with a puzzle in higher dimensions. As the bare S-matrix in higher dimensions is already IR finite, it is not apriori clear from this perspective whether one should be dressing the S-matrix or not. However the need for dressing can arises from an attempt to define gauge invariant observables in gravity \cite{Donnelly:2015hta}. It is proven in four dimensional spacetimes \cite{Choi:2017ylo, AtulBhatkar:2019vcb}  that the S-matrix which conserves the supertranslation charge is the KF dressed S-matrix and it is reasonable to expect that a similar proof should hold in higher dimensions.

 %Since the symplectic form defines the canonically conjugate variables in the phase space, we see that the canonically conjugate variable of $C_{ab}$ is proportional to $\int du \tilde D_{ab}$. And when promoted to the quantum theory, there is an uncertainty relation between two canonically conjugate variables. Since $\tilde D_{ab}$ defines the radiative degrees of freedom of the graviton, it is present in any quantum theory. And therefore, from the uncertainty principle, one simply cannot freeze the value of $C_{ab} = 0$ and thus is forced to have supertranslations, in order to get a sensible quantum description. 

%%%%%%%%%%%%%%%%%%%%%%%%%%%
\subsection{Summary} \label{summary}
In this paper we discuss supertranslations in (even) higher dimensions, specifically focusing on the six-dimensional case. We first specify the free data of the theory, which are given by the first two subleading coefficients in the large$-r$ expansion of $g_{ab}$. The graviton is defined by a combination of this free data ($C_{ab}$ and $D_{ab}$), which we call $\tilde D_{ab}$. This redefined field $\tilde D_{ab}$ contains the radiative data. The redefinition does not effect the News tensor ($\p_u \tilde D_{ab}$), but is necessary for the graviton to have the correct asymptotic fall offs and further, a finite symplectic form in the full non-linear theory. We emphasize that the redefinition is also required even if one is studying the linearized theory about a non-trivial Minkowski background ($C_{ab} \neq 0$). In the quantum theory, this implies that the graviton gets redefined depending on the vacuum one is working with. We discuss the symmetries of supertranslations in terms of the redefined variable $\tilde D_{ab}$.

We compute the supertranslation charge using the covariant phase space formalism. For this, we first evaluate the Symplectic form of the theory. We find that the symplectic form is uniquely split into two parts: one which is finite and characterizes the radiation of the system, and the other which is divergent (when integrated along $\scrip$). As the name suggests, the entire radiation content of the theory is contained in the finite \& radiative part. The split is unambiguous and the ambiguity is fixed by noting that only the radiative part leads to the correct Noether charge. In fact, upon trying to construct the charge from the divergent piece, we notice that it cannot be expressed as a total variation. Therefore, the radiative symplectic form at $\scrip$ is uniquely defined and it also helps us understand the canonically conjugate variables in the theory. As we point out, the symplectic form in the non-linear theory is a simple generalization of the result in \cite{Aggarwal:2018ilg}, and therefore the Noether charge can be obtained by following similar steps. We also compute the Bondi mass and evaluate the supertranslation charge using that. We find an exact matching between the two expressions -- the one via the radiative symplectic form and the one via the Bondi mass. The Noether charge in the full non-linear theory is a simple generalization of the result in the linearized theory about $C_{ab}= 0$. Upon combining with the results in \cite{Kapec:2015vwa} it is easy to see that we end up getting a similar Ward identity and therefore, the same structure for the Weinberg Soft theorem. In appendix \ref{app:QEDform}, we show how the gravitational case generalizes from the electromagnetic case which helps us understand these issues better.

Finally, we move onto a discussion of the memory effect and the IR triangle in the six dimensional non-linear theory. We find that the generalization works in a similar way as that of the Noether charge, where we simply have to replace $D_{ab}$ (in the linearized answers) with $\tilde D_{ab}$, to get the result in the non-linear theory. Therefore, this gives a very neat generalization of the IR triangle in higher-dimensions in the non-linear theory. As per our knowledge, this is the first example of the IR triangle in the non-linear theory in higher dimensions. In appendix \ref{app:memory} we also compare the final form of our answer with \cite{Pate:2017fgt} in the linearized limit and show that it is gauge invariant. Finally, in section \ref{sec:quantumdisc} we mention the important implications of our analysis in the quantum theory. 

%%%%%%%%%%%%%%%%%%%%%%%%%%%
\subsection{Future Directions} \label{futuredirections}
One of the major motivations for this analysis was to understand the principle of holography of information \cite{Raju:2019qjq, Laddha:2020kvp, Chowdhury:2020hse, Raju:2020smc, Chowdhury:2021nxw} in higher dimensional flat spacetime. This would be a direct extension of \cite{Laddha:2020kvp} and would require a proper understanding of the Hilbert space in higher dimensions. In this paper we have taken the first step of identifying the right phase space. It still remains to be explored as to what implications this has in the story of the principle of holography of information. We leave this for  future work.

%It would also be nice to understand how the asymptotic symmetries work near black hole horizons in higher dimensions \cite{Hawking:2016msc, Hawking:2016sgy, Dandekar:2016fvw}. 

The main focus of this paper has been the study of supertranslations in higher dimensions. It would be also be interesting to study  superrotations \cite{Colferai:2020rte}. Our analysis has been restricted to even dimensional spacetimes and we would like to explore the case of odd dimensions \cite{Fuentealba:2021yvo} in future.

%This paper initiates a study of these questions in the non-linear theory of gravity (or even in the linearized regime about a non-trivial vacuum) and we hope that this is useful in the study of many further questions in this field. 

\section*{Acknowledgments}
We are grateful to Alok Laddha for collaboration and many  discussions, for carefully reading the draft and providing suggestions. We also thank Suvrat Raju for suggesting the problem and for several discussions. We would also like to thank Anupam A. H, Sasank Budaraju, Jewel Kumar Ghosh, Victor Godet, Arpan Kundu, Prahar Mitra, Olga Papadoulaki, Priyadarshi Paul and Pushkal Shrivastava for several discussions. We have made use of the excellent software CADABRA \cite{Peeters2018} for several computations in this paper. We thank the organizers of LETHEP lecture series and the Indian Strings Meet 2021 for giving us an opportunity to present this work. RM would like to thank ICTS for the opportunity to participate in the LTVSP 2020 - 2021 programme and the INSPIRE fellowship for funding. CC and SGP acknowledge the support of the Department of Atomic Energy, Government of India, under project identification nos. RTI4001, RTI4002.

\begin{appendix}
%%%%%%%%%%%%%%%%%%%%%%%%%%%
\section{Metric Inverse and Traces}\label{app:inversetrace}
%%%%%%%%%%%%%%%%%%%%%%%%%%%
In this appendix we expand upon the computation of the metric inverse and the trace of various components in the metric. In the linearized theory this computation is fairly trivial but in the non-linear theory, it becomes slightly convoluted. The inverse metric $g^{ab}$ corresponding to the metric \eqref{bigmetric}, is given as
\be
g^{ab} g_{bc} = \delta^a_c . 
\ee
We will expand the LHS order by order in $r$ and then evaluate the $r$-expansion of the metric $g^{ab}$. Since $g_{bc}$'s leading order term is $r^2$, we expect that the leading order term of $g^{ab}$ will be $1/r^2$. Therefore, let us take the expansion of $g^{ab}$ as, 
\be
g^{ab} = \sum_{n = 2}^{\infty}\frac{g^{(n) ab}}{r^n}
\ee
and contract this with  
\be
g_{bc} = r^2 \delta_{bc} + r C_{bc} + D_{bc} + \frac{E_{bc}}{r} + \frac{F_{bc}}{r^2} + \cdots 
\ee
Noting that the indices of $g^{(n)}_{ab}$ are lowered and raised with $\gamma_{ab}$ we get the values of $g^{(n)}_{ab}$
\begin{align}
 g^{(2)ab} &= \delta^{ab}, \\
 g^{(3)ab} &= -C^{ab}, \\
 g^{(4)ab} &= C^a_c C^{bc} - D^{ab}, \\
 g^{(5)ab} &= C^{ac} D^b_c + D^{ac} C^b_c - C^a_m C^{mn} C^b_n - E^{ ab}, \\
 g^{(6)ab} &= - F^{ab} + C^{ac} E^b_c + C^{bc} E^a_c + D^{ac} D^b_c + C^a_m C^{mn} C^c_n C_{c}^b \\
 &\quad - (D^b_{c} C^a_m C^{cm} + D^c_{m} C^{am} C_{c}^b + D^a_{c} C^b_m C^{cm}) \nonumber.
\end{align}

%%%%%%%%%%%%%%%%%%%%%%%%%%%
\subsection*{Trace of $g^{(n)}_{ab}$}
We will now demonstrate how the Bondi gauge condition  \eqref{eq:detcondition} fixes the traces of $g_{ab}^{(n)}$.
The condition we have is $\det(\frac{g_{AB}}{r^2}) = \det(\delta_{ab})$. This condition results in, 
\begin{eqn}
 \det\left( \frac{g_{ab}}{r^2} \right)  &= \det\left( \delta_{ab} + \frac{C_{ab}}{r} + \frac{D_{ab}}{r^2} + \frac{E_{ab}}{r^3} + \frac{F_{ab}}{r^4} \right) \\
 &= \det (\delta_{ab}) \times \exp\tr\log \Bigl( \delta^c_b + \frac{C_b^c}{r} + \frac{D_b^c}{r^2} + \frac{E_b^{c}}{r^3} + \frac{F_b^{c}}{r^3}  \Bigr)
\end{eqn}
From the gauge condition \eqref{eq:detcondition} we see that we need, 
\be
\tr\log \Bigl( \delta^c_b + \frac{C_b^c}{r} + \frac{D_b^c}{r^2} + \frac{E_b^{c}}{r^3} + \frac{F_b^{c}}{r^3}  \Bigr) = 0. 
\label{trlogcond}
\ee
In order to simplify the term inside the $\tr$ we note that, 
\begin{eqn}
 &\log \Bigl( \delta^c_b + \frac{C_b^c}{r} + \frac{D_b^c}{r^2} + \frac{E_b^c}{r^3} + \frac{F_b^c}{r^4}   \Bigr)\\
 &= \frac{C_b^c}{r} + \frac{D_b^c - \frac{1}{2}C_b^m C_m^c}{r^2} + \frac{E_b^c - C_b^m D_m^c + \frac{1}{3} C_b^m C_m^n C_n^c}{r^3}\\
 &\quad + { F_b^c - C^{ab} E_{ab} - \frac{1}{2} D^{ab} D_{ab} + C^{am} C_{bm} D^b_a - \frac{1}{4} C^a_b C^b_c C^c_d C^d_a\over r^4} +O(1/r^5)
\end{eqn}
Taking the trace of this equation and using \eqref{trlogcond} then gives us \eqref{traceresults}. Although here we have shown how we can fix the traces of $g_{ab}^{(n)}$ for $n \leq 2$, this procedure is applicable $\forall \ n$.

%%%%%%%%%%%%%%%%%%%%%%%%%%%
\section{Equations of motion}\label{app:graveom}
%%%%%%%%%%%%%%%%%%%%%%%%%%%
 We eventually want to express everything in terms of free data, which, as argued in sec. \ref{sec:STshigher}  are $C_{ab}$ and $\tilde D_{ab}$. Therefore, using the Einstein equations we represent everything else in terms of those variables. We solve the Einstein equations in flat spacetime $R_{\mu\nu} = 0$ (where the specific fall off conditions are mentioned in \eqref{RicciFall}) and determine the components of the metric in terms of the free data. The ones which are important for us in this analysis are $R_{rr}$, $R_{ra}$, $R_{ur}$, $R_{ab}$, $R_{uu}$. We solve them order by order in $r$ in the large$-r$ limit. %As mentioned in sec. \ref{sec:STshigher} we will be working with $\gamma_{ab} = \delta_{ab}$, but the final results for the equations of motion will only change a little bit. 

 From the component $R_{ur}$ we have the following: 
 \begin{subequations}
 \be
 M^{(1)} = 0,
 \ee
 \be
 -M^{(2)} = \frac{1}{2} \p^a U^{(1)}_a +  \partial^2 \beta^{(2)} +   U^{(0)2}~. 
 \ee
 \label{Mneqns}
 \end{subequations}
 The next order equation does not yield us any non-trivial equation for $M^{(3)}$. For that one, we will have to work with the $R_{uu}$ equation which is like the Hamiltonian constraint.

 From the $R_{ra}$ component we obtain, 
 \begin{subequations}
  \be
  U_a^{(0)} = - \frac{1}{6} \p^b C_{ab}, 
  \ee
  \be
  3 U_a^{(1)} = - \p_b D^b_a + C_{ab} U^{(0) b} + \frac{1}{2} \p_b(C^{bm} C_{am}) + 6 \p_a \b^{(2)} + \frac{1}{8} \p_a(C^{bc} C_{bc}).
  \ee
    \label{eq:Uequation}
 \end{subequations}
 We can also explicitly write down the value of $U^{(2) a}$ but we will not be needing that for any specific calculation. All we need is the basic structure of $\partial_u U^{(2) a}$ for the computation of the Bondi mass evolution equation and we shall quote that here
 \begin{eqn}
  2 \p_u U^{(2) a} &= - \frac{3}{2} \p_u \p_b E^{ab} +\p_u \mathcal U(C, D) 
 \end{eqn}
 where $\mathcal U(C, D)$ is a bi-linear in $C_{ab}$ and $D_{ab}$. Using the fall off condition \eqref{Dtilfalloff} and the equation of motion for $E_{ab}$ (see equation \eqref{eq:Eabeom}), it is simple to see that $\int_{-\infty}^{\infty} du \ \p_u \p_a U^{(2) a} = 0$ and hence this does not contribute to the integrals in \eqref{Qst}.

From the $R_{rr}$ equation we obtain, 
\begin{subequations}
\begin{eqn}
\beta^{(2)} &= - \frac{1}{64}C^{ab}C_{ab}, \\
\beta^{(3)} &= \frac{1}{48} (C^{ab}C_{bc}C^c_a - 2 C^{ab} D_{ab}), \\
\beta^{(4)} &= \frac{1}{64}\Big[  5 C^{a}_b C^m_a D^b_m - 2 C^a_b C^b_c C^c_d C^d_a - 3 C^{ab} E_{ab} - D^{ab} D_{ab}  \Big].
\end{eqn}
\label{betaeqns}
\end{subequations}

In the linearized theory about $C_{ab} = 0$ it is clear that $\beta = 0$ \cite{Kapec:2015vwa, Aggarwal:2018ilg}. The equation of motion for $R_{ab}$ and $R_{uu}$ are discussed in the main text. To summarize them, the leading order non-trivial equation for $R_{ab}$ implies that $\p_u C_{ab} =0$ and the next non-trivial equation (at $O(1/r^2)$) gives us the value of $\p_u E_{ab}$ (see equation \eqref{eq:Eabeom}). There does not exist an equation of motion for $\p_u D_{ab}$ implying that is the free data of the theory. The leading non-trivial EOM for $R_{uu}$ gives the time evolution of the Bondi mass (see \eqref{mBevolution}).

%%%%%%%%%%%%%%%%%%%%%%%%%%%
\section{Memory in linearized gravity}\label{app:memory}
%%%%%%%%%%%%%%%%%%%%%%%%%%%
We can derive the memory in the linearized theory by using the formulas in \cite{Pate:2017fgt}. Even though the results in there are derived in the Harmonic gauge, the final answer is shown to match with ours (in the linearized limit), which reflects the fact that memory is gauge invariant (refer to section \ref{sec:memory}). Note that the results in \cite{Pate:2017fgt} are derived on the compact celestial sphere $\mathbb S^4$, whose curvature contributes to the final answer. To avoid a confusion with notations, we use the $\gamma_{\mathbb S}^{AB}$ and $\mathcal D_{\mathbb S}$ to denote the metric and the derivative on the compact sphere. The notations for the important metric fluctuations in \cite{Pate:2017fgt} are related to ours as, 
\bes
\bm{h}^{(0)}_{AB} \equiv D_{ab}, \qquad 
\bm h^{(1)}_{AB} \equiv E_{ab}
\ees
and the relative transverse displacement in the linearized case is denoted as $s_{lin}$. Therefore, using equation 3.4 and 3.5 \cite{Pate:2017fgt} of  we get, 
\be\label{slin}
\Delta s_{lin}^A = \frac{\gamma_{\mathbb S}^{AC}}{2r^3} \Delta \bm h^{(1)}_{CB} s_{lin,i}^B
\ee
Next we consider their equation 4.17 which gives (here the factor of $-4$ appears due to the curvature of the sphere), 
\be
\p_u \bm h_{BC}^{(1)} = - \frac{1}{2}(D_{\mathbb S}^2 - 4)\bm h_{BC}^{(0)}
\ee
Substituting this in \eqref{slin}, we get, 
\be
\Delta s_{lin}^A = - \frac{1}{4r^3} (D_{\mathbb S}^2 - 4)\int du \  \bm h^{(0)A}_{B} s_{lin,i}^B
\ee
Since $\bm h^{(0) AB} \equiv D^{ab}$ this expression can be obtained from \eqref{nonlinmem} in the linearized limit and upon taking care of the curvature of the sphere. 

%%%%%%%%%%%%%%%%%%%%%%%%%%%
\section{Lessons from QED Symplectic Form}\label{app:QEDform}
%%%%%%%%%%%%%%%%%%%%%%%%%%%
We demonstrate a detailed computation of the symplectic form in source-free QED as a toy model, and also describe how there are similar issues which crop up in gravity. Note that the behavior of linearized gravity is exactly similar to QED and hence we should expect the QED analysis to behave similarly to the linearized GR analysis done in \cite{Aggarwal:2018ilg}, with the small difference that we are working with $\gamma_{ab} = \delta_{ab}$. We encourage the reader to look at \cite{Freidel:2019ohg, Henneaux_2019} for a more mathematically robust treatment of the symplectic form in QED.

Before going to the computation of the symplectic form, let us mention the quantities that parametrize the phase space. The free data here is $A^{(0)}_i$ and $A^{(1)}_i$ and these are similar to $C_{ab}$ and $D_{ab}$ in gravity, as discussed in section \ref{sec:STshigher}. This can be shown by analyzing the Maxwell Equations. The notation for the fields in the large $r$ expansion here is the same as the one in gravity, 
\be
A_\mu(u, r, z) = \sum_{n} \frac{A_\mu^{(n)}(u, z)}{r^n}. 
\ee
We shall be working the radial gauge $A_r = 0$, which is analogous to the Bondi gauge in gravity \cite{Kapec:2014zla}. The other subsidiary conditions in this gauge choice are $A_u^{(0)} = A_u^{(1)} = 0$.

From the equations of motion we can show that
\be
\p_u A_i^{(0)} = 0
\label{Ai0eom}
\ee
which is analogous to \eqref{eq:Cab}. There does not exist an equation of motion for $A_i^{(1)}$ and that in general can be $u$-dependent, hence along with it being the free data, it is used to parametrize electromagnetic radiation. This  is analogous to $\tilde D_{ab}$ in gravity. 

We now proceed with the computation of the symplectic form. To do this, we first evaluate the symplectic current (whose general expression is given in \cite{Crnkovic:1986ex}). Like we did for the case of gravity, the component of the symplectic form that we are interested in $\mathcal  J^t$ (where $t = u + r$)
\be
\mathcal J^t = \mathcal J^u + \mathcal  J^r = - (\mathcal  J_r + \mathcal  J_u). 
\ee
The general form of $\mathcal  J_\a$ is given as \cite{Crnkovic:1986ex}, 
\be
\mathcal  J_\a = g^{\mu\nu}\delta A_\nu \wedge \d F_{\mu\a}. 
\ee
Thus we have to now compute the value of $\mathcal J_r$ and $\mathcal J_u$, which are given as, 
\be
\mathcal J_r =g^{\mu\nu} \delta A_\nu \wedge \delta F_{\mu r} = \frac{\delta^{ij}}{r^4} \delta A_j^{(0)} \wedge \delta A_i^{(1)}.
\label{eq:Jrem}
\ee
and 
\begin{eqn}
\mathcal J_u &= g^{\mu\nu} \delta A_\nu \wedge \delta F_{\mu u} = \frac{1}{r^2} \delta A^i \wedge \delta F_{iu}  \\
&= - \frac{1}{r^3} \delta A^{(0)i} \wedge \delta \p_u A_i^{(1)}  + \frac{1}{r^4} \Big[ - \delta A^{(1)i} \wedge \delta \p_u A_i^{(1)} + \delta A^{(0)i} \wedge \delta F_{iu}^{(2)} \Big]
\label{eq:Juem}
\end{eqn}
Where to get the $1/r^4$ term we have used $A_u^{(0)} = A_u^{(1)} = \p_u A_i^{(0)}= 0$. The first two are part of the gauge condition and the last one follows from an equation of motion \eqref{Ai0eom}. The currents can be simplified using the equation of motion which are given below. We specifically need the $\nabla_\mu F^{\mu r} = 0$ equation, and that is given as, 
\be
\p_u F_{ru} + \frac{1}{r^2} \p^i F_{iu} = 0 \implies 
\p^i F_{iu}^{(2)}= \p_u F_{ur}^{(4)}. 
\label{eq:eomem}
\ee
Where the second equation is one of the terms in the $r$-expansion of the first one. 
%%%%%%%%%%%%%%%%%%%%%%%%%%%
%%%%%%%%%%%%%%%%%%%%%%%%%%%
\subsection{Understanding $\mathcal J^t_{div}$}\label{sec:Jtdivem}
Next, we look at the $1/r^3$ term appearing in $\mathcal J_u$. This is analogous to the term $J^r_{div}$ appearing in  gravity (see \eqref{Jrdivgrav-1}). Naively this would lead to a divergent term in the expression for the symplectic form as the volume of spacetime comes with a determinant factor of $\sqrt{-g} = r^4$ which in the $r \to \infty$ limit will give a divergence with this. Therefore, we need to be a bit careful in order to handle this. This term in the current is labeled as $\mathcal J^t_{div} \equiv \mathcal J^u_{div}$ and is given as,
\be
\mathcal J_u^{div} = - \frac{1}{r^3} \delta A^{(0)i} \wedge \delta \p_u A_i^{(1)}
\label{Judivem}
\ee
In order to simplify this we use the fact that we are working with constant $t$-slices and we define $u = t - r$. By using this, we interpret the $1/r^3$ in the expression as $r/r^4$ and the $r$ on the numerator here will be written as $r = t - u$, where we hold $u$ constant as we take $r \to \infty$\footnote{These limits are in general hard to make sense of, and we do not provide a rigorous mathematical argument to prove this here. For a more robust mathematical discussion of the symplectic form in QED, we refer the reader to \cite{Freidel:2019ohg}.}. Therefore,
\begin{eqn}\label{JudivEM}
 \mathcal J_u^{div} &= - \frac{1}{r^3} \delta A^{(0)i} \wedge \delta \p_u A_i^{(1)} =  \frac{u - t}{r^4} \delta A^{(0)i} \wedge \delta \p_u A_i^{(1)}\\
 &\quad= \frac{1}{r^4} \p_u \Big[ (u - t) \delta A^{(0)i} \wedge \delta A_i^{(1)} \Big] - \frac{1}{r^4} \delta A^{(0)i} \wedge \delta A_i^{(1)} .
\end{eqn}
Notice that while computing the symplectic form we are eventually interested in $\int_{-\infty}^{\infty} \mathcal J_u^{div}$ and with the fall off
\be
\lim_{u \to \pm\infty}A_i^{(1)} \sim O(\frac{1}{|u|^{2 +0_+}})
\label{A1ifalloff}
\ee
we see that the first term in equation \eqref{JudivEM} vanishes (by repeating a similar analysis as that of appendix \ref{app:saddle} we also get this fall off using the saddle point approximation in QED). Therefore, we get, 
\be
\int_{-\infty}^{\infty} du  \ \mathcal J_u^{div} = - \frac{1}{r^4} \int_{-\infty}^{\infty} du \ \delta A^{(0)i} \wedge \delta A_i^{(1)}
\label{eq:Judivem}
\ee
With this in place, we simplify the full symplectic form, i.e, the combination of \eqref{eq:Jrem} and \eqref{eq:Juem}.

%%%%%%%%%%%%%%%%%%%%%%%%%%%
%%%%%%%%%%%%%%%%%%%%%%%%%%%
\subsection{Simplifying $\mathcal J^t$}
From the computation above, we see that upon combining  \eqref{eq:Jrem} and \eqref{eq:Judivem} we get, 
\be
\int_{-\infty}^{\infty} du \ \mathcal J_u^{div} +\mathcal  J_r = 0. 
\ee
In order to further simplify, we shall assume that we have no magnetic charges present (where the analogous statement in gravity is \eqref{eq:weylmagnetic-1}) and therefore the curl of $A_i^{(0)}$ is zero. Which means that $\ep^{ijk} \p_j A_k^{(0)} = 0 $ and this leads to the condition, 
\be
A_i^{(0)} = \p_i \Phi
\ee
where $\Phi(z)$ is some function on $\mathbb R^4$ and is like the ``potential for the vector potential $A_i$''. Like in the case of gravity, there is a freedom of adding a constant term in this expression, but that is not going to affect any of our analysis and hence we set that to zero. Therefore, in terms of $\Phi$, the symplectic form becomes, 
\begin{eqn}
\Omega^t_{QED}(\delta, \delta') = \int_{-\infty}^{\infty} du r^4 \mathcal J_u 
&= \int_{-\infty}^{\infty} du - \delta A^{(1)i}\wedge \delta' \p_u A^{(1)}_i - \delta  \Phi \wedge \delta' \p_u \p^i F_{ur}^{(4)} .
\end{eqn}
Where we have used the EOM in \eqref{eq:eomem} to get the second term. From this equation we can read of the poisson brackets of the free theory, and these are stated in \cite{Kapec:2014zla}. The first term in this like the kinetic term for the degree of freedom capturing radiation $A_i^{(1)}$. The second term gives us the conjugate to the soft mode. This is in contrast with the 4D case, as in there both of these come from the gauge field component at the same order $A_i^{(0)(4D)}$.

%%%%%%%%%%%%%%%%%%%%%%%%%%%
%%%%%%%%%%%%%%%%%%%%%%%%%%%
\subsection{Charge} \label{app:QEDcharge}
The charge is constructed using the variation under large gauge transformation. By working in the radial gauge we have the following variations,
\be
\delta_\ep A_i^{(0)} = \p_i \ep,\qquad 
\delta_\ep A_i^{(1)} = 0.
\ee
And therefore the charge is going to become,
\be
\delta Q_\ep =  \Omega^t(\delta_\ep, \delta)\implies Q_\ep = \int d^4 z\ \ep(z)  \big. F_{ur}^{(4)} \big|_{\mathcal I^-_+ } 
\ee
where we are neglecting the contribution at $\scrip_+$ due to the absence of massive particles. This is the same expression for the charge in QED as given in equation 3.2 of \cite{He:2019jjk}.

%%%%%%%%%%%%%%%%%%%%%%%%%%%
\section{Details of the Gravity Symplectic form}\label{app:gravform}
%%%%%%%%%%%%%%%%%%%%%%%%%%%
In this section we give a detailed description of the derivation of the symplectic form in gravity. Our starting point will be the Witten-Crnkovic current as given in \eqref{eq:gravform-1}. As stated before, from the determinant condition of the Bondi gauge \eqref{eq:detcondition} we have $\delta g = 0$ and hence the equation for the symplectic current simplifies to, 
\be
J^\alpha = \frac{1}{2}\delta \Gamma^\a_{\mu\nu}\wedge  \delta g^{\mu\nu}  - \frac{1}{2}\delta \Gamma_{\mu\nu}^{\nu} \wedge  \delta g^{\a \mu} 
\label{eq:gravform-2}
\ee
The components of interest in here will be $J^t$, which in turn just reduces to $J^r + J^u$. Therefore, we proceed onto evaluating these two. Since we are in 6-dimensions, the symplectic form constructed out of the current $J^t$ is integrated with the measure $r^4 \sqrt{\gamma} = r^4$ (when $\gamma_{ab} = \delta_{ab}$). Therefore we shall expand $J^t$ in inverse powers of $r$ and ignore the terms which are $O(1/r^5)$ and higher orders.

%%%%%%%%%%%%%%%%%%%%%%%%%%%
\subsection{$J^r$}
This is given as, 
\be
2J^r = \delta \Gamma^r_{\mu\nu}\wedge  \delta g^{\mu\nu}  - \delta \Gamma_{\mu\nu}^{\nu} \wedge  \delta g^{r \mu} 
\ee
By directly expanding this we get,
\begin{eqn}
-2J^r &= \delta g^{ru} \wedge (2 \Gamma^r_{ur} - \Gamma^\mu_{u \mu}) + \delta g^{rr} \wedge \delta( \Gamma^r_{rr} - \Gamma^\mu_{\mu r}) \\
&+ \delta g^{ra} \wedge \delta (2 \Gamma^r_{ra} - \Gamma^\mu_{\mu a}) + \delta g^{ab} \wedge \delta \Gamma^r_{ab}
\end{eqn}
We see that the terms in the first line are of higher order, i.e, 
\be
\delta g^{ru} \wedge (2 \Gamma^r_{ur} - \Gamma^\mu_{u \mu}) + \delta g^{rr} \wedge \delta( \Gamma^r_{rr} - \Gamma^\mu_{\mu r}) = O(1/r^5)
\ee
And thus, for getting the terms at the required order we have, 
\be
-2J^r =\delta g^{ra} \wedge \delta (2 \Gamma^r_{ra} - \Gamma^\mu_{\mu a}) + \delta g^{ab} \wedge \delta \Gamma^r_{ab}
\label{Jrfullgrav-1}
\ee

These terms are simple to compute and we list the final forms of the individual terms in here, 
\begin{enumerate}
 \item 
 \begin{eqn}
  \delta g^{ra} \wedge \delta (2 \Gamma^r_{ra} - \Gamma^\mu_{\mu a}) &= \frac{1}{r^4} \Big[ 2 \delta (U^{(1)a} - C^{ab} U_b^{(0)}) \wedge \delta U_a^{(0)} + \delta U_a^{(0)} \wedge \delta (3 U_a^{(1)} - C_{ac} U^{(0)c}) \Big]
  \label{Jrfullgrav-2}
 \end{eqn}

 \item 
 \begin{eqn}
  \delta g^{ab} \wedge \delta \Gamma^r_{ab}
  &= - \frac{1}{r^3} \delta C^{ab} \wedge \delta \Gamma^{r(0)}_{ab} + \frac{1}{r^4} \Big[ - \delta C^{ab} \wedge \delta \G^{(1) r}_{ab} + \delta(C^a_c C^{bc} - D^{ab}) \wedge \delta \G^{(0)r}_{ab} \Big]
  \label{Jrfullgrav-3}
 \end{eqn}

\end{enumerate}

Here $\Gamma^{(n)r}_{ab}$ denotes the coefficient of $1/r^n$ of $\Gamma^r_{ab}$ (which are given in \eqref{eq:GammarABcorrect}). Therefore we see that the terms in $J^r$ contain terms at $O(1/r^3)$ and $O(1/r^4)$. As we had mentioned before, in order to get the symplectic form, we need to integrate this with the measure which contains $r^4$ and therefore the term in $J^r$ containing $1/r^3$ has to be handled with caution. We shall treat these two separately below. This is treatment is analogous to the treatment of EM in \ref{sec:Jtdivem}. We will call the $1/r^3$ term $J^r_{div}$ and the $1/r^4$ term as $J^r_{fin}$.

%%%%%%%%%%%%%%%%%%%%%%%%%%%
\subsection{$J^r_{div}$}\label{app:Jrdiv}
Here we have,
\be
- 2J^r_{div} = - \frac{1}{r^3} \delta C^{ab} \wedge \delta \Gamma^{r(0)}_{ab} = - \frac{1}{r^3} \Big[  \delta C^{ab} \wedge \delta \p_{a} U_{b}^{(0)} + \frac{1}{2} \delta C^{ab} \wedge \delta \p_u D_{ab} \Big].
\label{Jrdivgrav-1}
\ee
The first term in this is actually zero when we look at its contribution in the symplectic form. And this can be seen by integrating on the flat sphere ($\gamma_{ab} = \delta_{ab}$) and using the equation of motion for $U_a^{(0)}$ \eqref{eq:Uequation},
\begin{eqn}
\int d^4 z \ \delta C^{ab} \wedge \delta \p_a U_b^{(0)} &\sim
\int d^4 z \ \delta C^{ab} \wedge \delta \p_a \p^c C_{bc}\\
&= \int d^4 z \ \p_a\Big[  \delta C^{ab} \wedge \delta  \p^c C_{bc} \Big] - \int d^4 z \ \delta \p_a   C^{ab} \wedge \delta  \p^c C_{bc} = 0. 
\end{eqn}
Here the first term is a total derivative on the flat sphere and hence is zero, and the second term is a wedge product of the same object and hence zero as well.

Let us look at the other term in \eqref{Jrdivgrav-1}. This term is similar to that of \eqref{Judivem} and thus can be handled in the same way. We shall also be using \eqref{Dtildedefn} and \eqref{eq:Cab} to write $\p_u D_{ab}= \p_u \tilde D_{ab}$. This replacement is done because, as motivated before,  $\tilde D_{ab}$ is the radiative data in the full non-linear theory. It will be demonstrated below as to how this is also necessary for defining a finite symplectic form and is also implied by the Saddle point approximation (see appendix \ref{app:saddle}). Therefore, the final form of \eqref{Jrdivgrav-1} becomes,
\be
-2 J^r_{div} = - \frac{1}{2r^3} \delta C^{ab} \wedge \delta \p_u \tilde D_{ab}
\ee
Like we did in the case of QED, we would like to think of the $1/r^3$ term as $r/r^4$ which is further thought of as $(t - u)/r^4$ where we would like to hold $t$ fixed as we take $r \to \infty$. Thus we get, 
\begin{eqn}
 - 2J^r_{div} = - \frac{1}{2r^4} \p_u \Big[ (t - u) \delta C^{ab} \wedge \delta \tilde D_{ab} \Big] - \frac{1}{2r^4} \delta C^{ab} \wedge \delta \tilde D_{ab} .
 \label{Jrdiv1}
\end{eqn}
We will describe how the finiteness of the symplectic form as computed from this expression fixes the fall off for $\tilde D_{ab}$.

%%%%%%%%%%%%%%%%%%%%%%%%%%%
\subsection{$J^r_{fin}$}
We now study the other part of $J^r$, which we had called $J^r_{fin}$. That is read off from \eqref{Jrfullgrav-1}-\eqref{Jrfullgrav-3}, 
\begin{eqn}
 2J^r_{fin} &= \frac{1}{r^4} \Big[  2 \delta (U^{(1)a} - C^{ab} U_b^{(0)}) \wedge \delta U_a^{(0)} + \delta U_a^{(0)} \wedge \delta (3 U_a^{(1)} - C_{ac} U^{(0)c}) \\
 &\qquad - \delta C^{ab} \wedge \delta \G^{(1) r}_{ab} + \delta(C^a_c C^{bc} - D^{ab}) \wedge \delta \G^{(0)r}_{ab}  \Big].
\end{eqn}
With a bit of work and integrating out total derivatives on the flat sphere, this becomes, 
\begin{eqn}
 - 2J^r_{fin}&=-\frac{1}{2} \delta D^{ab} \wedge \delta \p_u D_{ab}  - \frac{1}{2} \delta C^{ab} \wedge \delta \p_u E_{ab}\\
  &\quad-\frac{1}{2} \delta C^{ab} \wedge \delta \bigg[ 2 \p_{a} U_{b}^{(1)}  - U^{(0)c} \big( 2\p_a C_{bc}  - \p_c C_{ab} \big) - \frac{1}{3} \p_b\big[ U_a^{(1)} + C_a^c U_c^{(0)} \big] \\
  &\qquad \qquad \qquad\quad  + \frac{1}{3} \p_m \p_a (D_b^m - C_{c}^m C^{c}_b) \bigg] .
  \label{Jrfin1}
\end{eqn}
This equation can be simplified further and everything can be expressed in terms of $C_{ab}$ alone, which we will do after combining the results of $J^u$ and $J^r_{div}$.

%%%%%%%%%%%%%%%%%%%%%%%%%%%
\subsection{$J^u$}
From the general equation for the symplectic current in \eqref{eq:gravform-1} we have, 
\be
2J^u = \delta \Gamma^u_{\mu\nu} \wedge \delta g^{\mu\nu} - \delta \Gamma^\nu_{\mu\nu} \wedge \delta g^{u \mu}. 
\ee
This is relatively easy to compute and can be written as, 
\be
2J^u = \frac{1}{2 r^4} \delta C^{ab} \wedge \delta( C^a_c C^{bc} - D_{ab}) = - \frac{1}{2r^4}\delta C_{ab} \wedge \delta \tilde D^{ab} + \frac{3}{8r^4} \delta C_{ab} \wedge \delta (C^a_c C^{bc})
\label{Jugrav}
\ee
Where we have used $C^a_a = 0$.

%%%%%%%%%%%%%%%%%%%%%%%%%%%
\subsection{$J^t$}\label{app:Jtfull}
From the definition $t = u + r$, $J^t$ is given as $J^u + J^r$ and therefore combining \eqref{Jrdiv1}, \eqref{Jrfin1} and \eqref{Jugrav}, we have, 
\begin{eqn}
 2J^t &= -\frac{1}{2 r^4} \delta D^{ab} \wedge \delta \p_u D_{ab}  - \frac{1}{2 r^4} \delta C^{ab} \wedge \delta \p_u E_{ab} - \frac{1}{2r^4} \p_u \Big[ (t - u) \delta C^{ab} \wedge \tilde D_{ab}\Big] \\
  &\quad-\frac{1}{2} \delta C^{ab} \wedge \delta \bigg[ 2 \p_{a} U_{b}^{(1)}  - U^{(0)c} \big( 2\p_a C_{bc}  - \p_c C_{ab} \big) - \frac{1}{3} \p_b\big[ U_a^{(1)} + C_a^c U_c^{(0)} \big] \\
  &\qquad \qquad \qquad\quad  + \frac{1}{3} \p_m \p_a (D_b^m - C_{c}^m C^{c}_b) + \frac{1}{4} C^c_a C_{bc} \bigg] .
  \label{Jtfull-1}
\end{eqn}
From the equations of motion we have in sec.(\ref{app:graveom}) we can write $J^t$ completely in terms of $C_{ab}$ and $\tilde D_{ab}$. And after doing that, we will find three kinds of terms in the expansion. One of which is a remnant of $J^r_{div}$ and in the equation above that appears with a $\p_u \Big[ \cdots \Big]$. And in the other two terms, one of them contains the terms dependent on $\tilde D_{ab}$ and the other is independent of $\tilde D_{ab}$. These three parts are called $J^t_{div}$, $J^t_{fin}$ and $J^t_{NI}$ respectively. Here $J^t_{NI}$ stands for ``non-integrable'' and we will explain the meaning of that in more detail in the upcoming section.

%%%%%%%%%%%%%%%%%%%%%%%%%%%
\subsubsection{$J^t_{div}$}\label{app:gravJtdiv}
Here, 
\be
- 2J^t_{div}=- \frac{1}{2r^4} \p_u \Big[ (t - u) \delta C^{ab} \wedge \delta \tilde D_{ab} \Big]
\ee
This term will give a contribution to the symplectic form which looks like, 
\be
\Omega^t_{div}\sim \int du \ \p_u \Big[ (t - u) \delta C^{ab} \wedge \delta \tilde D_{ab} \Big] =  \Big[ (t - u) \delta C^{ab} \wedge \delta \tilde D_{ab} \Big]_{u \to \infty} - \Big[ (t - u) \delta C^{ab} \wedge \delta \tilde D_{ab} \Big]_{u \to - \infty} . 
\ee
The way in which we regulate this is to choose the fall off for $\tilde D_{ab}$ as, 
\be
\lim_{u \to \pm\infty}\tilde D_{ab} \sim O\Big( \frac{1}{|u|^{2 + 0_+}} \Big),
\label{tildeDfalloff-1}
\ee
which is exactly similar to the way in which $A_i^{(1)}$ falls off in EM which is discussed around \eqref{A1ifalloff}, and is the same fall-off as derived using the saddle point approximation in appendix \ref{app:saddle}. With this in place, we see that the contribution
to
\be
\Omega^t_{div} = 0.
\ee

Henceforth we shall be making the following replacement in the symplectic current, as it does not contribute to the form as motivated by the falloff \eqref{tildeDfalloff-1}, 
\be
\frac{1}{r^4} \delta C^{ab} \wedge \delta \p_u D_{ab} = \frac{1}{r^4} \delta C^{ab} \wedge \delta \p_u \tilde D_{ab} \to 0.
\label{replacementDDtilde}
\ee

%%%%%%%%%%%%%%%%%%%%%%%%%%%
\subsubsection{$J^t_{I}$}
After substituting the EOM in \eqref{Jtfull-1} we collect the terms dependent on $\tilde D_{ab}$ and that is equal to,
\be
-2J^t_{I} = - \frac{1}{2} \delta \tilde D_{ab} \wedge \delta\p_u \tilde D^{ab} + \frac{1}{2} \delta C^{ab} \wedge \delta \p_u E_{ab}+ \frac{1}{9} \delta C^{ab} \wedge \delta  \p_a \p^c \tilde D_{bc}. 
\ee
In order to get this we have used the fall off \eqref{tildeDfalloff-1} to replace, 
\be
\delta D^{ab} \wedge \delta \p_u D_{ab} = \delta \tilde D^{ab} \wedge \delta \p_u \tilde D_{ab}
\ee
which is possible because of the replacement \eqref{replacementDDtilde}. We are also including the $\p_u E_{ab}$ term as it is a function of $\tilde D_{ab}$ (see \eqref{eq:Eabeom}). It is interesting to note that the contribution to the symplectic form due to this term is the same as the linearized case (about $C = 0$ vacuum as in \cite{Aggarwal:2018ilg}), but with $D_{ab}$ replaced by $\tilde D_{ab}$. 

Therefore, the integrable part of the symplectic form becomes, 
\be
\Omega^t_{I} = \intI d^4 z du J^t_{I} = - \frac{1}{2}\intI d^4 z du - \frac{1}{2} \delta \tilde D_{ab} \wedge \delta\p_u \tilde D^{ab} + \frac{1}{2} \delta C^{ab} \wedge \delta \p_u E_{ab}+ \frac{1}{9} \delta C^{ab} \wedge \delta  \p_a \p^c \tilde D_{bc}. 
\ee
It has been shown in section \ref{sec:symplecticform} that this leads to the correct soft charge, which is obtained by convoluting $m_B$ with a function $f(z)$ (see \eqref{Qst}).

%%%%%%%%%%%%%%%%%%%%%%%%%%%
\subsubsection{$J^t_{NI}$}
Let us consider the final piece of the symplectic current \eqref{Jtfull-1} now, which solely depends on $C_{ab}$.  Given the form of \eqref{Jtfull-1}, a tedious but straightforward computation leads us to,
\begin{eqn}
 -2J^t_{NI} =\delta C^{ab} \wedge &\delta \Big[-\frac{1}{2} C_a^c C_{bc}-\frac{1}{9} \p_a \Big(  C^{c d} \p_c C_{b d} + \frac{4}{3} C_b^c \p^d C_{c d} - \frac{1}{16} C^{cd} \p_b C_{cd}  \Big) \\
 &\quad + \frac{1}{3} \p^c C_{cd} \p_a C^d_b - \frac{1}{6}\p^c C_{ab} \p^d C_{cd} \Big].
\end{eqn}

%%%%%%%%%%%%%%%%%%%%%%%%%%%
\section{Saddle Point analysis}\label{app:saddle}
%%%%%%%%%%%%%%%%%%%%%%%%%%%
We perform the saddle point approximation for the Graviton to get an idea of the fall-off of the field at $\mathcal I^+_\pm$. Since we are only interested in the fall-offs, we will be cavalier about the numerical constants as they will not be necessary for the final answer. We will follow the treatment in \cite{Strominger:2017zoo}\footnote{Exercise 4 in \cite{Strominger:2017zoo} treats the 4D electromagnetic case, but the formalism is very similar.}.

The mode expansion of the Gravitational perturbation in Cartesian coordinates is given as
\be
h_{\mu\nu}(u, \vec r) = \sum_\alpha \int \frac{d^5 q}{(2\pi)^5 2 |\vec q|}\Big[   \ep_{\mu\nu}^{(\alpha)} a_{\alpha}(|\vec q|, \hat q) e^{i q \cdot x}  + h.c. \Big]
\ee
Here $\a$ is the polarization index, $\ep_{\mu\nu}^{(\a)}$ represents the polarization vector and $a_{\a}$ represents the mode functions. The factor $e^{i q \cdot x}$ is given as, 
\be
e^{i q \cdot x} = e^{- i q t + i \vec q \cdot \vec x} = e^{- i q u - i q r (1 - \cos \theta)},
\ee
where $\theta$ is the angle between $\vec x$ and $\vec q$ and we often use $|\vec q| \equiv q$ when there is no conflict of notation. With this expansion in place, we evaluate the integral in the $r \to \infty$ limit. Along with the $e^{i q \cdot x}$ term there is another contribution of $\theta$ that comes from the measure of the integral, which is proportional to $\sin^3\theta$ (since it is a Five Dimensional integral). Thus the necessary contribution of the integral becomes, 
\be
h_{\mu\nu}(u, \vec r) \sim \sum_{\alpha} \int dq d \theta \ q^3 \sin^{3}\theta  \big[ \ep^{(\a)}_{\mu\nu} a_{\a}   e^{- i q u - i q r (1 - \cos \theta)}  + h.c. \big]
\ee
Now we consider the saddle point approximation of the integral in the $r \to \infty$ limit. Here the saddle we pick is $\theta = 0$, where the other one is forbidden by the Riemann-Lebesgue lemma. The leading order saddle point answer will give us a $0$, hence we have to go the subleading order, which means that we consider the following expansions, 
\be
\lim_{r \to \infty} \sin^{3}\theta e^{-i q r(1 - \cos\theta)} \approx  \lim_{r \to \infty} \theta^{3} e^{-\frac{i q r \theta^2}{2}}
\ee
After which the integral is proportional to,
\be
\lim_{r \to \infty }h_{\mu\nu} \sim \sum_{\alpha} \int dq d\theta \ \theta^{3} q^{3} \big[ \ep^{(\alpha)}_{\mu\nu} a_{\a}   e^{- i q u - \frac{i q r \theta^2}{2}}  + h.c. \big]
\ee
Since we only want the $r$ and $q$ dependence of the $\theta$ integral we get, 
\be
\lim_{r \to \infty }h_{\mu\nu} \sim \frac{1}{r^2}\sum_{\alpha} \int dq \frac{q^{3}}{q^{2}} \Big[   \ep^{(\alpha)}_{\mu\nu} a_{\alpha}  e^{- i q u } k_1+ h.c. \Big]= \frac{1}{r^2}\sum_{\alpha} \int dq q\Big[   \ep^{(\alpha)}_{\mu\nu} a_{\alpha}  e^{- i q u } k_1 + h.c. \Big].
\ee
where $k_1$ is an unimportant constant used to represent the numerical value for the integral over $\theta$. In order to get the behavior of the Graviton field at $\mathcal I^+_{\pm}$ we now need to take the limit $|u| \to \infty$. Again, via saddle point approximation, this will now enforce the limit $q \to 0$ on the integrand of RHS. Using the normalization of the soft factor as given in \cite{Laddha:2019yaj} (see equations 2.10 and 2.29 of the reference mentioned) we get the behavior of the mode functions at $q \to 0$ in six Dimensions
\be
\lim_{q \to 0}\ep_{\mu\nu}^{(\a)} a_\a (q, \hat q) \propto q^0 k_2(\hat q)
\ee
where $k_2(\hat q)$ is an unimportant function depending on the direction of $\hat q$. This leads us to the asymptotic fall off at $\mathcal I^+_{\pm}$of $h_{\mu\nu}$ 
\begin{eqn}
 \lim_{u \to \pm \infty}\lim_{r \to \infty }h_{\mu\nu} \sim \frac{1}{r^2}\lim_{q \to 0} \int dq q e^{i q u} k_1  k_2(\hat q) = \frac{1}{r^2} \times  O\Big( \frac{1}{u^2} \Big)
\end{eqn}
where the final step can be seen by performing a simple change of variable $q u = l$ for example. Since the physical Graviton mode is given as $\tilde D_{ab}$ (see eq. \eqref{Dtildedefn}) we can relate $h_{\mu\nu}$ to $D_{ab}$ by a simple coordinate transformation
\be
\tilde D_{ab} = \lim_{r \to \infty}\frac{\p x^\mu}{\p x^a} \frac{\p x^\nu}{\p x^b} h_{\mu\nu} \propto  \lim_{r \to \infty} r^2 h_{\mu\nu}
\ee
and thus we arrive at the fall-off for the graviton field near $\mathcal I^+_{\pm}$ (see \eqref{Dtilfalloff})
\be
\lim_{u \to \pm\infty}\tilde D_{ab} = O\Big( \frac{1}{u^2} \Big).
\ee
A similar analysis can also be performed for the Photon in six dimensions and we end up with a similar fall-off condition for the radiative mode in QED.

\end{appendix}

\typeout{}
%\bibliography{references}

\printbibliography %Prints bibliography

\end{document}